\documentclass[mathleft]{an}
\usepackage{amssymb}
\usepackage{amsmath}
\usepackage{multirow}
\usepackage{longtable}
\usepackage{threeparttable}
\usepackage{graphicx}
\usepackage{times}

\def\enerunit{erg~sec$^{-1}$~cm$^{-2}$ }
\def\etal{et al.}

\def\Journal#1#2#3#4{: #3, {#1} #2, #4}
\def\AA{A.\&A.}
\def\ANA{AN}
\def\APJ{ApJ.}
\def\APS{ApJS.}

\def\AST{Astron.J.}

\def\JCL{J.Classification}

\def\MRA{MNRAS}

\def\PAS{Publ.Astron.Soc.Pac.}

\def\be{\begin{equation}}
\def\ee{\end{equation}}
\def\bea{\begin{eqnarray}}
\def\eea{\end{eqnarray}}

\begin{document}
\title{Statistical Algorithms for Identification of Astronomical X-Ray Sources}

\author{Houri Ziaeepour\inst{1}\fnmsep\thanks{Corresponding author:
  \email{hz@mssl.ucl.ac.uk}\newline}
\and Simon Rosen\inst{1}}

\titlerunning{Statistical Algorithms for Identification}
\authorrunning{H. Ziaeepour \& S. Rosen}
\institute{Mullard Space Science Laboratory, Holmbury St. Mary, Dorking, 
Surrey RH5 6NT, UK.}

\received{----}
\accepted{----}
\publonline{later}
\keywords{X-rays: general --- methods: statistical --- surveys}

\abstract
{Observations of present and future X-ray telescopes include a large number of 
serendipidious sources of unknown types. They are a rich source of knowledge 
about X-ray dominated astronomical objects, their distribution, and their 
evolution. The large number of these sources does not permit their individual 
spectroscopical follow-up and classification. Here we use Chandra 
Multi-Wavelength public data to investigate a number of statistical algorithms 
for classification of X-ray sources with optical imaging follow-up. We show 
that up to statistical uncertainties, each class of X-ray sources has specific 
photometric characteristics that can be used for its classification. 
We assess the relative and absolute performance of classification methods and 
measured features 
by comparing the behaviour of physical quantities for statistically classified 
objects with what is obtained from spectroscopy. We find that among methods 
we have studied, multi-dimensional probability distribution is the best for 
both classifying source type and redshift, but it needs a sufficiently large 
input (learning) data set. In absence of such data, a mixture of various 
methods can give a better final result. We discuss some of potential 
applications of the statistical classification and the enhancement of 
information obtained in this way. We also assess the effect of classification 
methods and input data set on the astronomical conclusions such as 
distribution and properties of X-ray selected sources.}

\maketitle

\section{Introduction}
Since the launch of Chandra and XMM-Newton X-ray observatories at the end of 
90's our vision of the X-ray sky has been tremendously changed. The large 
collecting area of the XMM-Newton, and the sensitivity and high spatial 
resolution of Chandra (\cite{cxcres}) have increased our knowledge about the 
details of the morphology and the energy spectrum of extended sources and the 
distribution and nature of point sources. They have also made it possible to 
study fainter and harder sources, and related physical processes. Each 
observed field, even with relatively modest exposure time of few kilo seconds, 
adds tens of new objects to the list of serendipitous sources. This gives the 
opportunity to increase our knowledge of individual sources as well as the 
general characteristics of classes of astronomical objects, and thereby the 
modeling of the underlying physical phenomena. In addition, in a large set 
of objects with various characteristics and behaviour, there is always the 
chance to find new classes/subclasses of sources or extreme and rare examples 
of known types (\cite {faintx};\cite {faintx1}). 

Multi-wavelength spectroscopic follow-up of X-ray selected sources based on 
Chandra (\cite {champ1}; \cite {champ3}) and 
XMM-Newton (\cite{axis}; \cite {axis0}; \cite {roberto}; \cite {brightsam}) 
source catalogs 
have already investigated a number of important astronomical issues such as 
redshift evolution of different classes of Active Galactic Nuclei (AGN) 
(\cite{agndust}; \cite {agndust1}; \cite {agndust2}; \cite {agndust3}; 
\cite {agndust4}; 
\cite {agndust5}), their intrinsic absorption (\cite{agnabs}; \cite{agnabs1}; 
\cite {agnabs2}; \cite {agnabs3}) and their relationship with the 
characteristics of the host galaxy, its star formation history 
(\cite{agnstar}; \cite {agnstar1}; \cite {agnstar2}; \cite {agnstar3}; 
\cite {agnstar4}; \cite {agnstar5}) and the central super-massive black hole 
(\cite{agnbh}; \cite {agnbh1}; \cite {agnbh2}). Observations 
have also revealed the relation between hard X-ray sources, hidden broad line 
AGN(BLAGN) or BL-Lac(\cite{agnhard}; \cite {agnhard1}; \cite {agnhard2}) 
and X-ray background in hard (\cite{agnxcb}; \cite {agnxcb1}; 
\cite {agnxcb2}; \cite{agnxcb3}; \cite {agnxcb4}) and soft bands 
(\cite{softxcb}). 

Although these observations have significantly improv-ed our knowledge about 
X-ray sources (\cite{bepphard}; \cite {bepphard1}; \cite {bepphard2}), they 
only include relatively small number of 
spectroscopically identified X-ray objects. Despite the fact that they belong 
to randomly selected fields, it is not certain how much other criteria 
such as brightness of the optical counterparts and the tendency to study 
special classes of sources more than others, influence conclusions 
about populations and their physical conditions. Thus, it seems  
that much larger and unbiased samples representing different category of 
objects, with both typical characteristics of the population, and from 
extreme parts of the observable parameter space are necessary. It is 
however very difficult to obtain spectroscopic data for such large samples 
in a reasonable time. 

An alternative to spectroscopic identification is classification based on 
multi-wavelength, photometric quantities such as fluxes, count-rates and 
hardness ratios. Abstractly speaking, the identification analysis exploits 
characteristics of the parameter space of spectroscopically identified objects 
to establish a series of rules to distinguish different classes. Unclassified 
objects are associated to a category according to these rules. This is similar 
to spectroscopic identification where observer must know the main spectral 
features discriminating one category of sources from others. The knowledge 
about characteristics of each class is either provided by the observer from 
prior information or can be found or learned in an automatic way using already 
classified objects. Some of the popular learning algorithms (\cite{algo}) are 
mapping (\cite{mapping}), neural network (\cite{neuralnet}; 
\cite {neurnetapp}), nearest neighbour (\cite{algo}; \cite {algo1}) and 
statistical algorithms (\cite{stat})(Citations are examples and are not 
exclusive). These classification algorithms have an 
intrinsically statistical nature and as we will see later, depend on how much 
the learning sample is statistically representative of the population. For 
this reason we generally call them statistical identification methods.

In the present work we describe a number of statistical methods for 
classifying X-ray selected sources with candidate optical counterparts. Our 
aim is not just suggesting few algorithms but also assessing their reliability 
and their effect on astronomical conclusions. The input set of 
spectroscopically classified sources are taken from the publicly available 
Chandra Multi-wavelength project (ChaMP) (\cite{champ2}) identifications. The 
unclassified sources to which we apply these methods are also taken from 
fields observed by ChaMP, and therefore are subject to the 
same selection conditions as the input data. No other prior assumption is made 
about sources or classes. As this paper concentrates on the technical 
aspects of classification, we don't investigate the completeness of the input 
sample, and therefore astronomical conclusions obtained from this data can 
not be extended to other data sets without caution.

In the following sections we first briefly review the X-ray and optical 
quantities used in the classification. Then, we explain methods we 
have investigated and compare their performance. Finally, we apply methods 
considered to be more efficient to a list of unclassified sources and show the 
effect of the input set as well as the classification method on the 
distribution of statistically classified sources and there-by, 
on the scientific conclusions.

\section {X-ray and Optical Data} \label {sec:data}
ChaMP includes 137 high galactic latitude fields selected for having hydrogen 
column densities $N_h < 6 \times 10^{20} cm^{-2}$ or optical extinction 
$E(B-V) < 0.1$mag (\cite{champ1}). For the first public release of ChaMP 
spectroscopical identifications, 62 of these fields which have PI 
authorization have been used. With additional constraints explained in detail 
in \cite{champ1}, it is expected that full ChaMP fields include 
$\sim 6000$ background X-ray sources and indeed the public data available 
from ChaMP web site\footnote{{\scriptsize http://hea-www.cfa.harvard.edu/CHAMP/IMAGES\_DATA/champ\_xpc.tab}} includes 6512 X-ray sources with photometric information i.e. 
count rates, fluxes and hardness ratios.
 
The X-ray flux is calculated in 5 overlapping energy bands defined in 
Table \ref{tab:xband}.

\begin{table}[ht]
\caption{ChaMP X-ray energy bands \label{tab:xband}}
\vspace{0.2cm}
\begin{tabular}{ll}
\hline Band & Energy \\
\hline Broad (B) & 0.3-8.0 keV \\
Hard (H) & 2.5-8.0 keV \\
Soft (S) & 0.3-2.5 keV \\
Soft1 (S1) & 0.3-0.9 keV \\
Soft2 (S2) & 0.9-2.5 keV \\
\hline
\end{tabular}
\end{table}
In the classification methods described here we use three distinct bands 
S1, S2 
and H. The broad band B is also used for normalisation when use, as parameter 
space, the ratio of fluxes to the X-ray flux. We also define two hardness 
ratios:
\bea
H_{S1S2} &=& \frac {S2-S1}{S2+S1}. \label {hss}\\
H_{S2H} &=& \frac {H-S2}{H+S2}. \label {hhs}
\eea
where $H$, $S1$ and $S2$ are total counts in the corresponding band. Energy 
conversion factors for 62 fields of the first release are given 
in \cite{champ1} for three power-law spectrum models with 
$\Gamma = 1.2,~1.4,~1.7$. For compatibility with what is used for the 
XMM-Newton data, through out this work we use fluxes determined with 
$\Gamma = 1.7$. 

Optical follow-up of the first 6 ChaMP fields for which observations and 
identifications are complete, was perform-ed by using NOAO telescope for both 
southern and northern fields (\cite{champ2}). Optical filters for imaging 
correspond to Sloan 
filters $i$, $r$ and $g$. For northern fields if the counterparts are not 
too faint, it is also possible to use Sloan public data. The advantage is that 
it includes 2 more filters $u$ and $z$. We will show later the importance of 
having information in more filters, especially in $u$. For this reason, we 
have 
retrieved the 5- filter optical magnitudes from the Sloan 4$^{th}$ release for 
northern fields. Between unclassified sources in ChaMP fields, we use only 
objects with all Sloan 5 filters. It is not however possible to apply the same 
restriction to spectroscopically identified sources, because their number is 
quite limited. For identified sources out of Sloan coverage, we use ChaMP 
published magnitudes and estimate the missing $u$ and $z$ magnitudes by 
modeling color difference for each category of sources. 

To model magnitudes, we define a color curve for each source with all the 5 
magnitudes mentioned above. They are ordered in increasing central frequency 
and the difference of successive magnitudes - colors - are determined. For 
each category of sources we obtain the average color curve, considered as 
template for the category, and its standard deviation. We assume a Gaussian 
distribution around this template. 

\begin{figure*}[t]
\begin{center}
\begin{tabular}{cc}
a) \includegraphics[height=7.5cm]{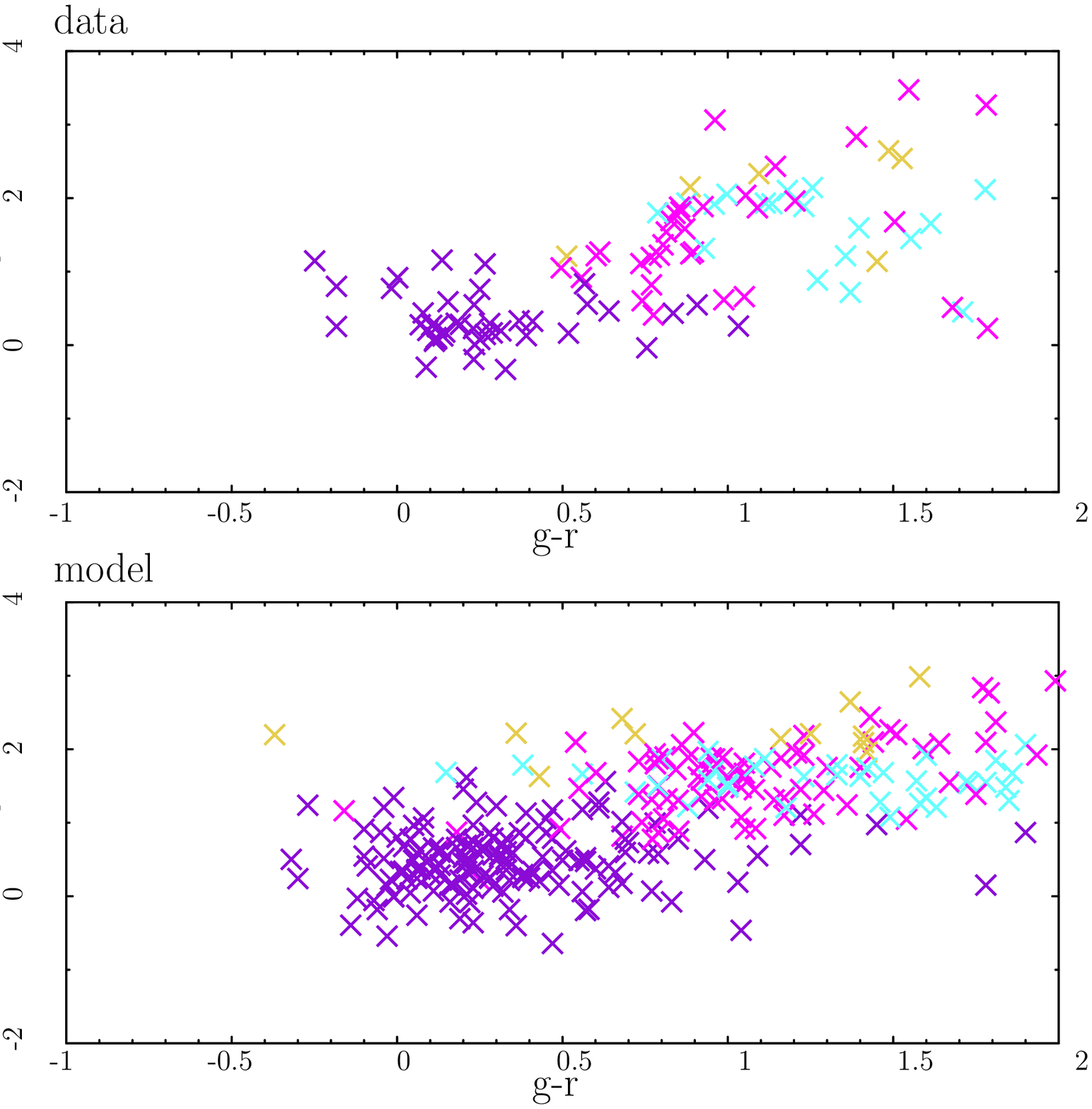} & 
b) \includegraphics[height=7.5cm]{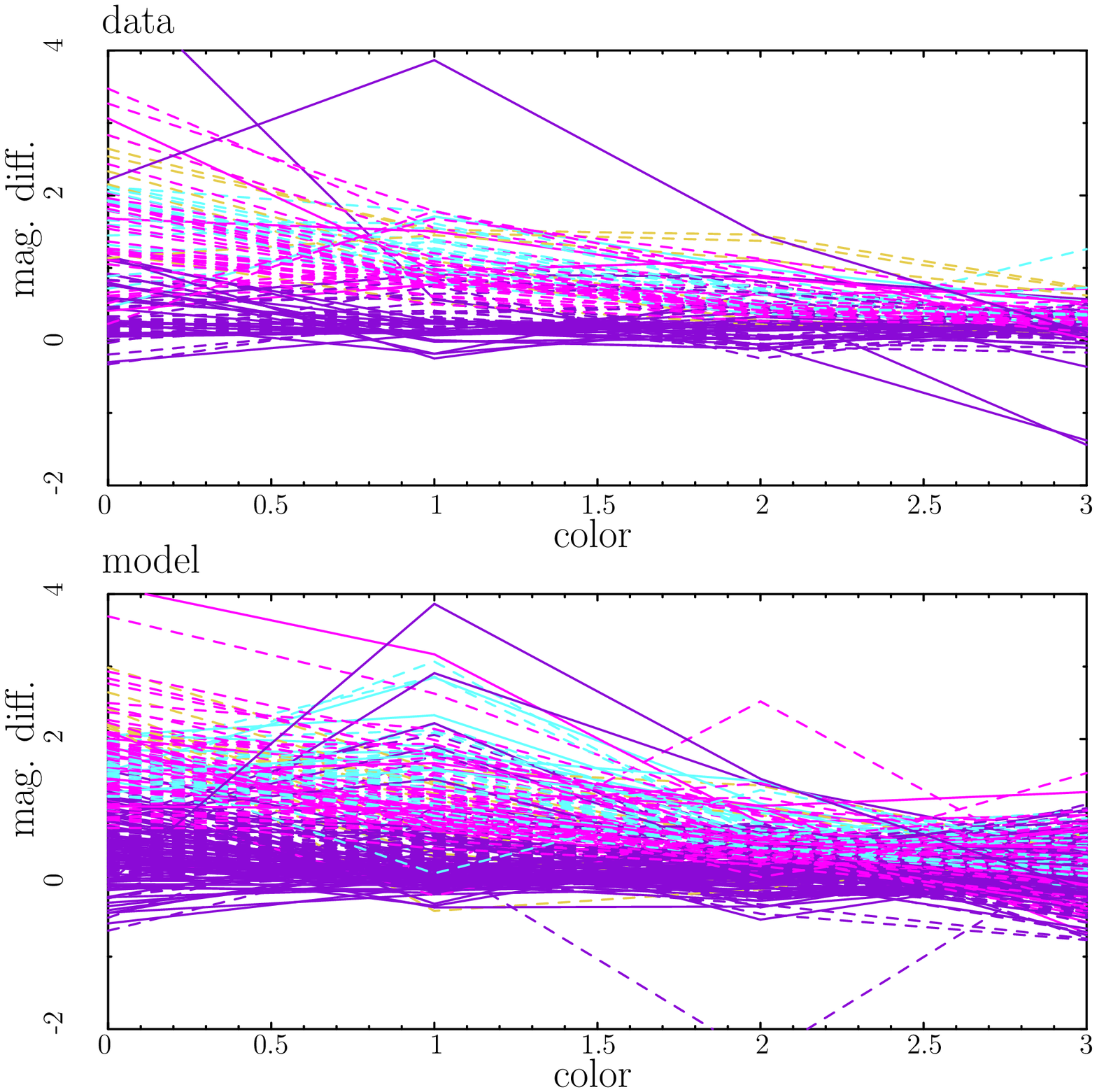} \\ 
\multicolumn{2}{c}{c) \includegraphics[height=7.5cm]{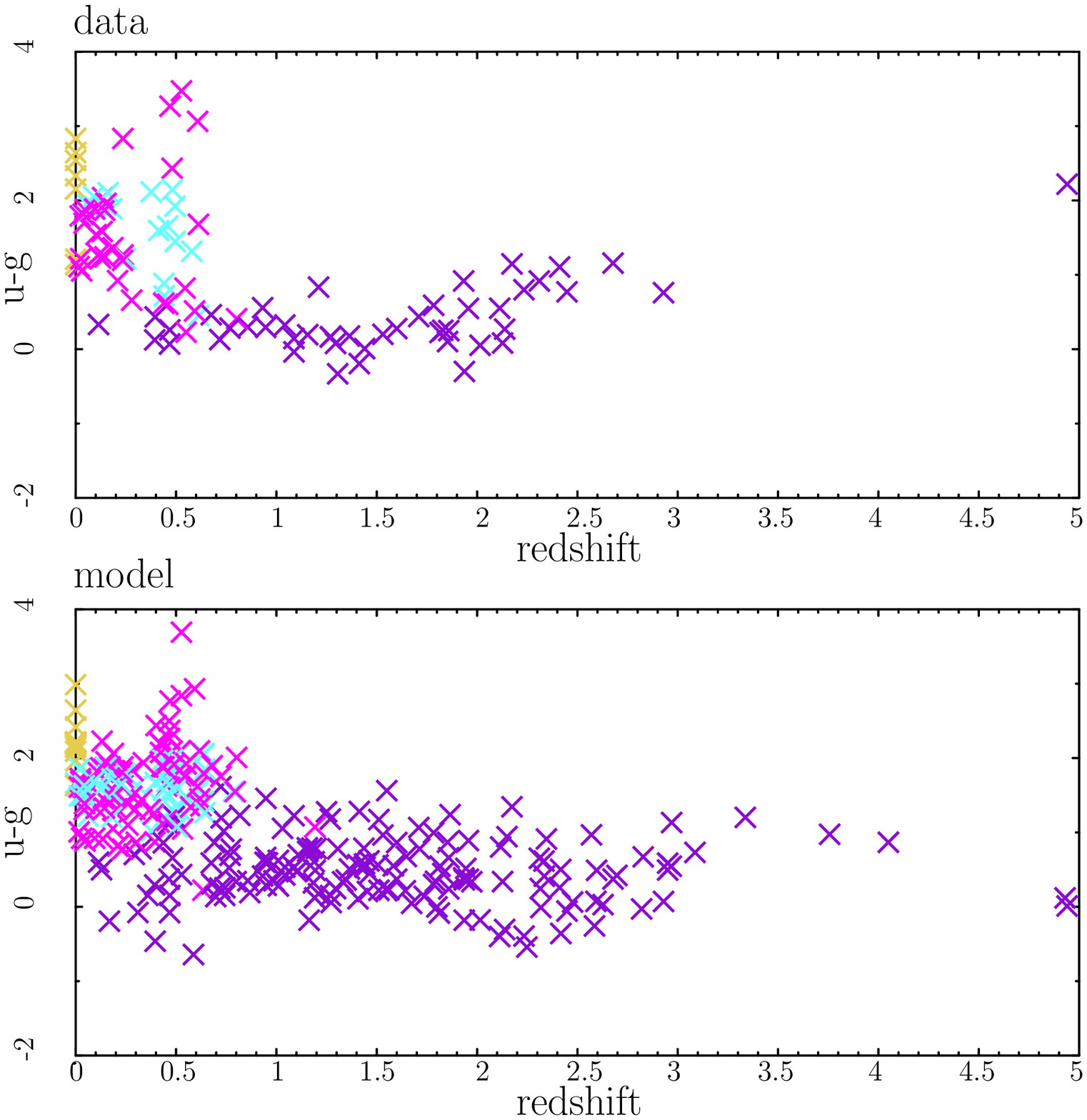}} 
\end {tabular}
\caption{(a) $u-g$ versus $g-r$ for data with $u$ magnitude (top) and modeled 
data without archival $u$ magnitude. Source type: BLAGN (violet), 
NLEG (magenta), ALG (cyan), star (gold). (b) Color distribution for sources 
in (a). Dash and full lines present respectively sources with redshift 
$\leqslant 1.5$ if the source is a BLAGN or $\leqslant 0.6$ otherwise. 
(c) Redshift distribution. To make the input sample to identification methods 
more uniform, 2 sources with largest deviation are removed before further 
application.
\label {fig:colormodel}}
\end{center}
\end{figure*}

To estimate $u$ and $z$ magnitude for identified sources without these 
measurements, we choose a random value for $u-g$ and $i-z$ according to a 
Gaussian distribution with average and deviation taken from the 
template color curve and its deviation. Figs. \ref{fig:colormodel}-a, b and c 
show the color and redshift distribution of sources with $u$ and $z$ 
magnitudes 
from Sloan archive and the distribution of sources for which their $u$ and $z$ 
magnitudes are estimated. These plots show that measured and modeled 
distributions are statistically similar. Evidently, at source by source level 
there is no guarantee that the estimated magnitudes correspond to real ones 
for a given source. But statistically speaking, there can be sources with 
$i$, $r$, and $g$ magnitudes 
as one of these sources and $u$ and $z$ magnitudes close to what we have 
estimated~\footnote{Note that we have used only optical magnitudes $i$, $r$, 
and $g$ for extrapolating to $z$ and $u$. It would be better to use both 
optical and X-ray fluxes, but this makes the model too complicated.}. 
This level of precision is enough for us because our purpose here is studying 
the methodology of classification and not real properties of the sources we 
use. Therefore, a set of data with statistically the same properties as what 
can be encountered in reality is sufficient.

\subsection {Spectroscopic Identifications}
The first ChaMP public data release includes 125 spectroscopically identified 
sources (\cite{champ2}). Most of them are in the medium flux category i.e. 
have $10^{-15} \lesssim f \lesssim 10^{-14}$ \enerunit 
in energy range $0.5$ keV $\leqslant E \leqslant 2$ keV. A small set of 
sources have fluxes as small as few times $\sim 10^{-16}$ \enerunit or as 
large as few times $\sim 10^{-13}$ \enerunit. Objects are classified in 4 
categories: Broad Line Active Galaxy Nucleus (BLAGN), Narrow Emission Line 
Galaxy (NELG), Absorption Line Galaxy (ALG) and star. The details of criteria 
for associating a source to one of these categories can be found in 
\cite{champ2}. For the purpose of completeness here we summarize them. BLAGN 
category is designated to sources with board band - 
FWHM $> 1000$ km~sec$^{-1}$ - emission lines. If all emission lines have 
FWHM $< 1000$ km~sec$^{-1}$, the source is classified as NELG. Objects with 
high $S/N > 20$ and no identifying feature are classified as BL-Lac. Sources 
with absorption lines are classified as ALG or stars. Their spectrum is 
compared to synthetic templates for classification and for determination of 
their redshift. A classification confidence flag is given to each 
source: 0 not classifiable, 1 insecure, 2 secure. Here we use only sources 
with secure identification in the input dataset.

The set of 125 serendipitiously identified sources is too small for 
statistical 
applications. Therefore, we have added also sources from two other published 
set of identified sour-ces by the ChaMP, i.e. hard AGNs (\cite{cxcagnhardnew}) 
and normal galaxies (\cite{cxcnormgal}). For this set of sourc-es as before we 
use the publicly available ChaMP data for x-ray, Sloan 
optical magnitude in 5 bands as explained above, and modeling when some of the 
optical magnitudes are mis-sing. After neglecting sources for which reliable 
data was not available and modeling couldn't give reasonable values, we put 
together a total of 268 sources with identifications and all the flux/magnitude 
information. This set includes 151 BLAGNs, 65 NELGs. 37 ALGs and 15 stars.
It is clear that in this set the number of stars is under-estimated with 
respect 
to what one can expect from a serendipidious selection of X-ray sources in 
high galactic latitude fields. Although the addition of ChaMP hard AGN and 
normal galaxies have increased the size of the learning set for the statistical 
identification, due to the bias of these set toward a special type of 
objects, the final set does not anymore presents what one can find in a set of 
randomly selected sources subjected only to conditions on X-ray fluxes and 
optical magnitudes. This is not a major problem for present work where the 
main purpose is the investigation of performance of classification methods. 
However, we will show later that the learning (input) set to algorithms or 
classifying agents such as humans, inevitably leaves its trace on the 
classification and this is inherent to the concept of learning and extension of 
information. This issue should be considered when the results of statistical 
classification are used for obtaining scientific conclusions.

\section{Statistical Classification} \label{sec:stat}
The first step in mining X-ray selected sources is their classification. In 
this work we concentrate on statistical algorithms rather than automatic 
class/cluster finding algorithms such as neural network. The main reason for 
this choice is that we are searching for classes of astronomical objects 
which are already defined. If one of these classes happens to cluster in the 
available parameter space with other objects of different astronomical 
categories, we are not discovering a new class or property but encountering 
obstructions to object classification. Cluster finding methods are certainly 
interesting when the internal structure of the data is unknown and blind 
clustering helps to organize objects with similar properties to groups. This 
makes their studying easier.

\begin{figure*}[t]
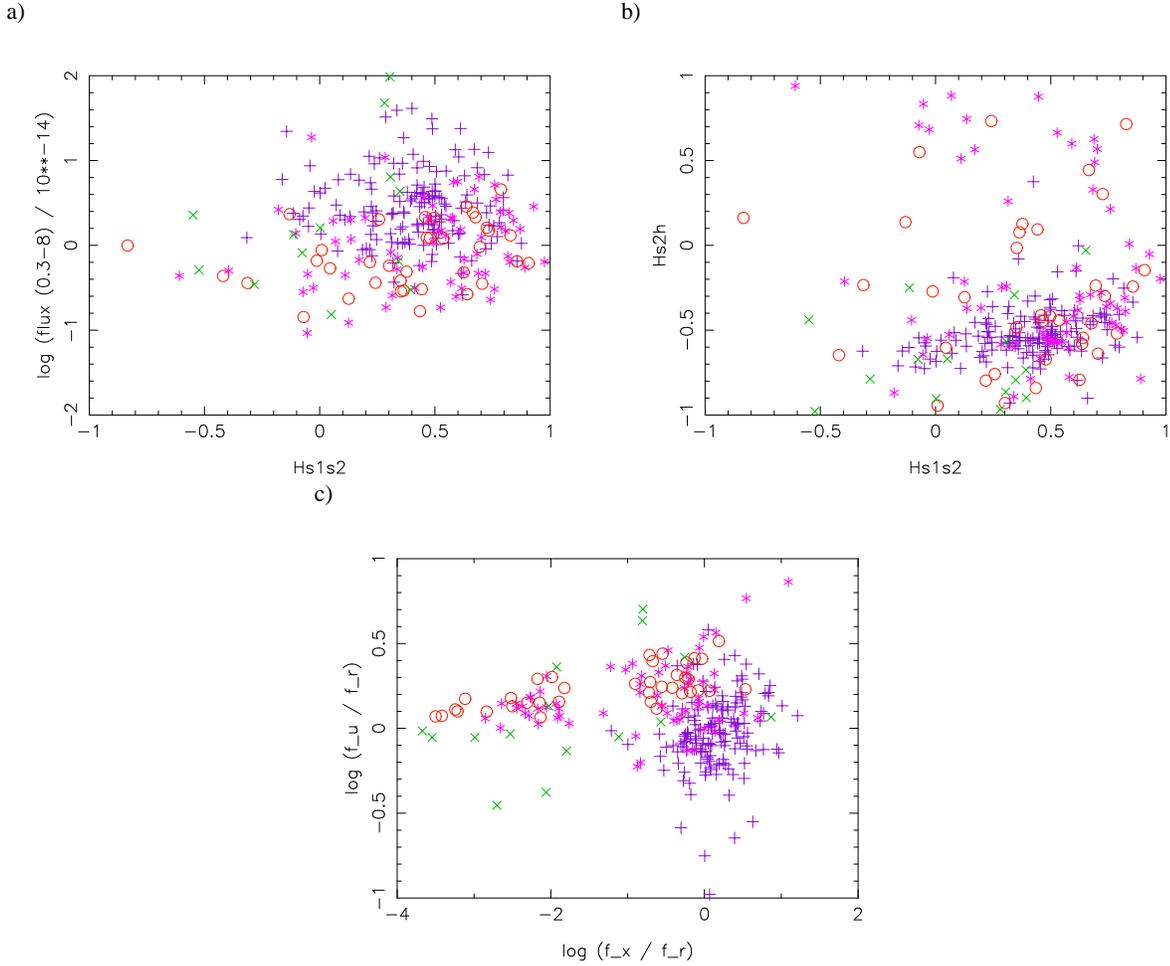

\begin{center}
\begin{tabular}{cc}
a)\includegraphics[height=7.5cm,angle=-90]{fig2a.eps} & 
b)\includegraphics[height=7.5cm,angle=-90]{fig2b.eps} \\ 
\multicolumn{2}{c}{c)\includegraphics[height=7.5cm,angle=-90]{fig2c.eps}} 
\end {tabular}
\caption{Distribution of spectroscopically identified ChaMP sources in 
$H_{S1S2}$- $0.3$ keV to $8$ keV flux (a), $H_{S1S2}$-$H_{S2h}$ (b) and 
$\log (f_x/f_r)-\log(f_u/f_r)$ (c) planes. X-ray flux $f_x$ is the same 
as in (a). $f_u$ and $f_r$ are optical fluxes in the corresponding bands.
BLAGN ($+$, Violet), NELG (star, magenta), ALG/Gal (circle, red), 
star ($\times$, green).}
\label{fig:hf}
\end{center}
\end{figure*}

Up to now attempts for statistical classification of X-ray sources were 
mostly concentrated on the relation between hardness ratios, fluxes and count 
rates (\cite{roberto}; \cite {xcolor}). As an example, Fig. \ref{fig:hf}-a and b 
show X-ray flux in ChaMP B band and hardness ratio $H_{S2h}$ versus hardness 
ratio $H_{S1S2}$ for ChaMP identified sources described above. It is clear 
that there is a large overlap between parameters of different classes of 
sources, and therefore the addition of other information like optical/IR data 
is necessary for a more reliable classification (\cite{houriproc}). Recent 
attempts for statistical classification of ROSAT sources 
 (\cite{rosatclass}; \cite {rosatradio}) also show a better classification 
efficiency when both X-ray and optical 
data are used. On the other hand, 2-dimensional parameter-space does not seem 
to be adequate. For instance, spectroscopic follow-up of Chandra and 
XMM-Newton sources and previous observations, show that in the 
medium range of X-ray fluxes and optical/IR magnitudes, most of the X-ray 
selected sources are BLAGN. They are mostly concentrated in a small area 
of $g-r$ and $u-g$ plane of optical data (Fig. \ref {fig:colormodel}-a), 
or flux and hardness ratio (Fig. \ref{fig:hf}-a), $H_{S1S2}-H_{S2h}$ 
(Fig. \ref{fig:hf}-b) planes of X-ray data or even in 
$\log (f_x/f_r)-\log(f_u/f_r)$ (Fig. \ref{fig:hf}-c) when both X-ray and 
optical data are used. This area however contains other and rarer types of 
sources, and on each of these planes  alone there is no significant separation 
of all the classes. This means that what ever the method of classification, 
the performance of a 2-dimensional parameter space would be 
low (\cite{houriproc}).

The alternative is using a multi-dimensional parameter space. In the next 
sections we explain a few methods of statistical classification which 
simultaneously explore all the available parameters. None of these methods are 
new, but for each application it is necessary to select the best parameters to 
use, the optimal number of parameters, and the performance of each method or 
their application in a complementary way (see below). This is the main aim of 
the present work. 

Although using more parameters for classification leads to a greater chance 
for having reasonably separate clusters of objects in the parameters space, 
each corresponding to one category, the larger number of parameters also 
expands the volume to fill with the input data. Therefore, adding new 
parameters sometimes deteriorates the classification performance. Moreover, 
computational limitations, both on the amount of memory necessary for 
buffering the parameter space and on the CPU time for calculation, constrain 
the optimal parameter set, range of parameters, and their binning.

The reason for studying multiple classification models is that each of them 
have its own advantages and drawbacks. The choice of a method depends on the 
input data-set available and the application of classification results. 
Two main categories of applications are short-listing X-ray selected sources 
for further follow-up, and studying physical properties of each class. In the 
first case a simple but mildly precise method and a small set of input data is 
adequate. For the second application a much better performance and therefore a 
much larger input set is necessary. For instance, as we will see in more 
details below, classification based on the 
distance to a cluster of objects is not very precise. However, it needs a much 
smaller input (learning) set than other methods. It can be applied when only 
a small set of identifications is available, and one needs to have a crude 
classification of other sources for further investigation - for instance when 
only one type of source is targeted. Multiple methods can be also used as 
complementary to each other. As an example, in the following sections we will 
explain in detail a multi-dimensional probability distribution which is the 
best method because it uses the full parameter space and correlation between 
various parameters, but it needs a large input set. As a consequence, when 
this condition is not fulfilled, many sources can not be classified and this 
has an important impact on astronomically important properties such as 
redshift distribution. We show that in this case using less precise but also 
less demanding methods to classify these remaining sources leads to a better 
overall performance. 

\subsection {Performance Estimation}
The usual practice for testing the performance of classification is to divide 
the set of known sources into two sets and use one for learning and the other 
for testing. At present however, our sample of identified sources is very 
small and dividing it to two parts certainly reduces the performance of the 
classification significantly and hides the real ability of methods
\footnote{For comparison it is worth to mention that the classification of the 
ROSAT sources in \cite{rosatclass} uses a selection tree method. When 
both X-ray and optical data are used their input set consist of 6763 RASS and 
9247 WGACAT sources. When only X-ray data is used, the size of the input 
data is yet larger. As mentioned above we have only 268 identified sources 
for which all the necessary optical and X-ray data is available.}. 
Note also that for a statically significant assessment of the performance, 
the test set must be also enough large. We can not use just a small subset of 
the identified sources for test. The test set must be enough large such that 
good or bad performance of the classification be statistically significant. In 
particular, if the input set is biased, the result of the test with a data set 
with very different distribution can be completely misleading. For instance, 
we know that in our ChaMP dataset stars are under-represented. If from this 
set we select a small subset of sources randomly, it can include one or two 
stars and the correctness of their subsequent classification is statistically 
insufficient to assess the quality of the classification method. 

The issue of the optimal minimum size of the input set is not well understood. 
The suggested value in \cite{tempclass}; \cite {tempclass1} is a minimum of 
10 times the number of 
categories in the classification\footnote{10 is simply the sample length 
considered to be minimal statistically significant sample for most statistical 
applications.}. In practice, the size of the input set must be much larger. In 
our case we want to classify X-ray selected 
sources to 4 types and 12 redshift ranges (see below for details). According 
to this prescription we need at least 480 spectroscopically identified 
sources. At
present we have only 268 ! Nonetheless, we show that even with such a small
set we can see the difference between the performance of the methods, and when
they are applied to a set of $\sim 1300$ unknown sources - much larger than
the input set - the distribution obtained for the physical quantities such as
$\log N-\log S$ or redshift distribution are generally what we expect. 

We try various strategies to assess the performance of the methods and 
the effect of having to use a small set of input data. 
We divide the set of spectroscopically identified sources to two sets by 
randomly selecting each source to be member of one or the other subset. We have
used an equal probability and also $60\%-40\%$ and $70\%-30\%$. We use the
larger set as input and the smaller one as test set. To see the performance 
in the case where a much larger input set is available, we also test 
algorithms by using the same set both as input and as test set.  In this case 
for some methods such as low resolution spectrum template fitting, the 
classification is $100\%$ correct, and therefore this does not permit 
a performance test. But for methods 
based on probability distributions, this is simply similar to the case where 
the distribution is known (see also below for more explanation). To use the 
largest possible input set and largest possible test set, we have also tried 
another procedure~\footnote{We thank M. Page for suggesting this procedure.}. 
In each run, we remove one of the sources from the set of spectroscopically 
identified sources, use the rest as input set and classify the removed 
source. We repeat the same procedure for all the members of the set and 
assess the performance by adding the result of all the iterations. This 
procedure needs much longer CPU time because at each iteration all the 
distributions must be recalculated. For this reason we apply it only to a 
limited number of cases. We find that this more complex test does not add much 
information about the performance of the method than what we conclude from 
simple division of available identified sources, but it helps to see that a 
larger input set definitively has an important impact on the classification 
performance. 

For quantifying the performance of classification methods 3 quantities have 
been suggested by \cite{classqual1}; \cite {classqual2}:
\bea
d_i &=& \text{Detection rate of class $i$} \nonumber \\
&=& \frac{\text{Number of objects statistically classified as $i$}}
{\text{Number of class i objects in the test set}}. \label{detectr} \\
r_i &=& \text{Reliability rate for class $i$} \nonumber \\
&=& \frac{\text{Number of 
correctly classified object}}{\text{Number of objects statistically 
classified as $i$}}. \label{relir}\\
c_{ij} &=& \text{Contamination rate of class j in class i} \nonumber \\
&=& \frac{\text{Num. of class j objects statistically classified as i}}
{\text{Number of objects statistically classified as i}}. \nonumber \\ 
& &
\label{contr}
\eea
Note that $c_{ij}$ defined in (\ref{contr}) is not the same as 
{\it false alarm rate} defined by \cite{classqual1}; \cite {classqual2}. 
They use these quantities to assess the classification IRAS sources as 
galaxy and non-galaxy (star). Here we have mor 
than just 2 categories, and the rate of contamination as defined in 
(\ref{contr}) is a better description of the performance of descrimination 
between classes. In fact the false alarm rate corresponds to the contamination 
rate of galaxies in non-galaxies defined in (\ref{contr}).

To assess the effect of imprecise identifications on the astronomical 
conclusions, we use redshift distribution of cl-asses as a benchmark. This is 
in fact one of the most important outcomes of the classification of a large 
number of X-ray selected sources. The main producers of the astronomical 
X-ray are accretion to black holes, star formation activities and galaxy 
clusters. 
Redshift distribution of these objects not only permits to study the 
evolution of star formation epochs, but also can help to better understand 
the issue of relation between the growing rate of super-massive black holes, 
AGN activities, and star formation. This has been the subject of intensive 
study in recent years. For stars this test is evidently irrelevant.

To quantify the closeness of distributions from spectroscopy and from 
statistical identification, we use a $\chi^2$ clo-seness test. For each class 
of objects we calculate a $\chi^2$-like quantity as the following:
\bea
\chi^2 = \frac{1}{Z - 1} \sum_{i=1}^Z \frac{(n_i - N_i)^2}{N_i^2}. 
\label {ztest}
\eea
where $n_i$ is the number of sources from statistical classification in that 
category and in $i^{th}$ redshift bin, $N_i$ is the same number from 
spectroscopy, $Z$ is the number of redshift bins with $N_i > 0$. If two 
distributions are exactly the same, $\chi^2 = 0$, otherwise it can be 
interpreted as an average difference or goodness of fit between two 
distributions. We also perform a Kolmogorov-Smirnov (KS) test of the redshift 
distribution~\footnote{We thanks R. Della Ceca for suggesting a KS 
test.}. In general, the conclusion of both tests about the 
performance is the same. The interpretation of $\chi^2$ test is however more 
direct and relevant for the interpretation of physical properties. The 
results of both tests along with quality rates defined in 
(\ref{detectr})-(\ref{contr}) are presented in Table \ref{tab:kstest}. 

We want to add another remark about the relevance of these tests as a means 
for performance assessment specially regarding the astronomical applications 
of the results. KS and $\chi^2$ tests here provide a rough estimation of how 
reliably the redshift distribution can be recovered by using statistical 
algorithms. Each of them is just one number summing and smearing all the 
features of a distribution. In an astronomical applications however features 
can have special meaning and interest. In comparing two distributions, one 
can have smaller $\chi^2$ and KS, but for instance have a peak or trough non 
existent in the real distribution. They can mislead the astronomical 
interpretation of distributions. Moreover, large number of 
unclassified sources can lead to a redshift distribution quite different from 
the original one. Nonetheless, the method can be more precise when a source 
gets a 
classification. This is specially the case for the last method we explain in 
the next section. In conclusion, the performance assessment based on fit or 
statistical methods are incomplete and this issue must be considered in the 
selection of methods for astronomical applications.

\section {Algorithms}
In this section we describe in details classification methods we have 
investigated and compare their performance. They can be summarized as the 
followings:
\begin{description}
\item{$1-$} {\bf Distance to Clusters:} Each class is characterized by a 
vector presenting the typical place of the objects in the parameter space and 
a deviation matrix. Unknown objects are classified according to their distance 
to each class.
\item{$2-$} {\bf Low resolution spectrum (photometry):} Fluxes are used 
to define a low resolution spectrum. Unknown sou-rces are classified both for 
their type and their redshift by fitting their spectrum to templates. This 
method is essentially the traditional photometric redshift determination to 
which we have also added the classification of sources.
\item{$3-$} {\bf Parameters as independent measures:} Each measured 
quantity is considered as an independent identifier of a class. For each class 
its distribution is determined, and Maximum Likelihood (ML) is used to 
estimate the likelihood for an object to belong to a class and a redshift.
\item{$4-$} {\bf Multi-dimensional probability:} The whole parameter space is 
used for the classification by binning parameters and calculating 
multi-dimensional probability distribution for each class. Unknown sources are 
classified according to the conditional probability to belong to one of the 
classes.
\end{description}
The results of all classification methods and their tests are summarized in 
Table \ref{tab:kstest}. 

\subsection {Distance to Clusters} \label{sec:proj}
This method is a variant of the clustering algorithm (\cite{stat}) in which an 
object is classified with respect to its distance to clusters of known 
sources. Clusters are defined by a vector presenting the center of the 
cluster and a deviation (covariance) matrix:
\bea
\vec {X}^i &=& \frac{1}{N^i}\sum_n \vec{x}^i_n. \nonumber \\
C^i_{\alpha\beta} &=& \frac{1}{N^i}\sum_n (x_{n\alpha} - X^i_{\alpha}) 
(x_{n\beta} - X^i_{\beta}). \label{dist} \\
&& n = 1 \ldots N^i \quad , \quad \alpha~,~\beta = 1 \ldots D. \nonumber
\eea
where $\vec{X}^i$ and $C^i_{\alpha\beta}$ are respectively center and 
deviation matrix of $i^{th}$ class, $D$ and $N$ are respectively the dimension 
of the parameter space and the number of input sources of type $i$. To 
classify an unknown source, its deviation from the center of each cluster is 
calculated, and it is associated to the closest cluster/class if its 
distance from the center of the cluster is smaller than the deviation of the 
class, otherwise it is marked as unclassifiable. The performance of this 
method is moderate, see Table \ref{tab:kstest}. Parameter space that has 
been used for this method and others is also defined in this table.

Another variation of cluster method is calculating distances to clusters 
projected to two subspaces of parameters. For the $i^{th}$ object:
\bea
D_{i1} &=& \biggl (\sum_n x_{in}^2 \biggr)^{\frac{1}{2}} \quad n = 1 
\ldots N_1. \nonumber \\
D_{i2} &=& \biggl (\sum_n x_{in}^2 \biggr)^{\frac{1}{2}} \quad n = N_1 + 1 
\ldots N. \label{distdim}
\eea
where $N_1$ is the dimension of the first subspace and $N_2 = N - N_1$ is the 
dimension of the second subspace. Distribution of known sources in the 
2-dimensional space $D_1-D_2$ is used for classification.

A natural choice for subspaces in our case is optical and X-ray parameters. 
After binning parameters, in each subspace we determine the Euclidean distance 
(\ref{dist}) with respect to the origin - the bin with 
smallest index (i.e. zero) for all the parameters. Normalization can be 
performed by choosing the same number of bins for all the parameters, 
otherwise the effect of parameters with more bins would be more than others. 
This also can be considered as a weight given to parameters which are more 
important in distinguishing one class from others.

\begin{figure*}[ht]
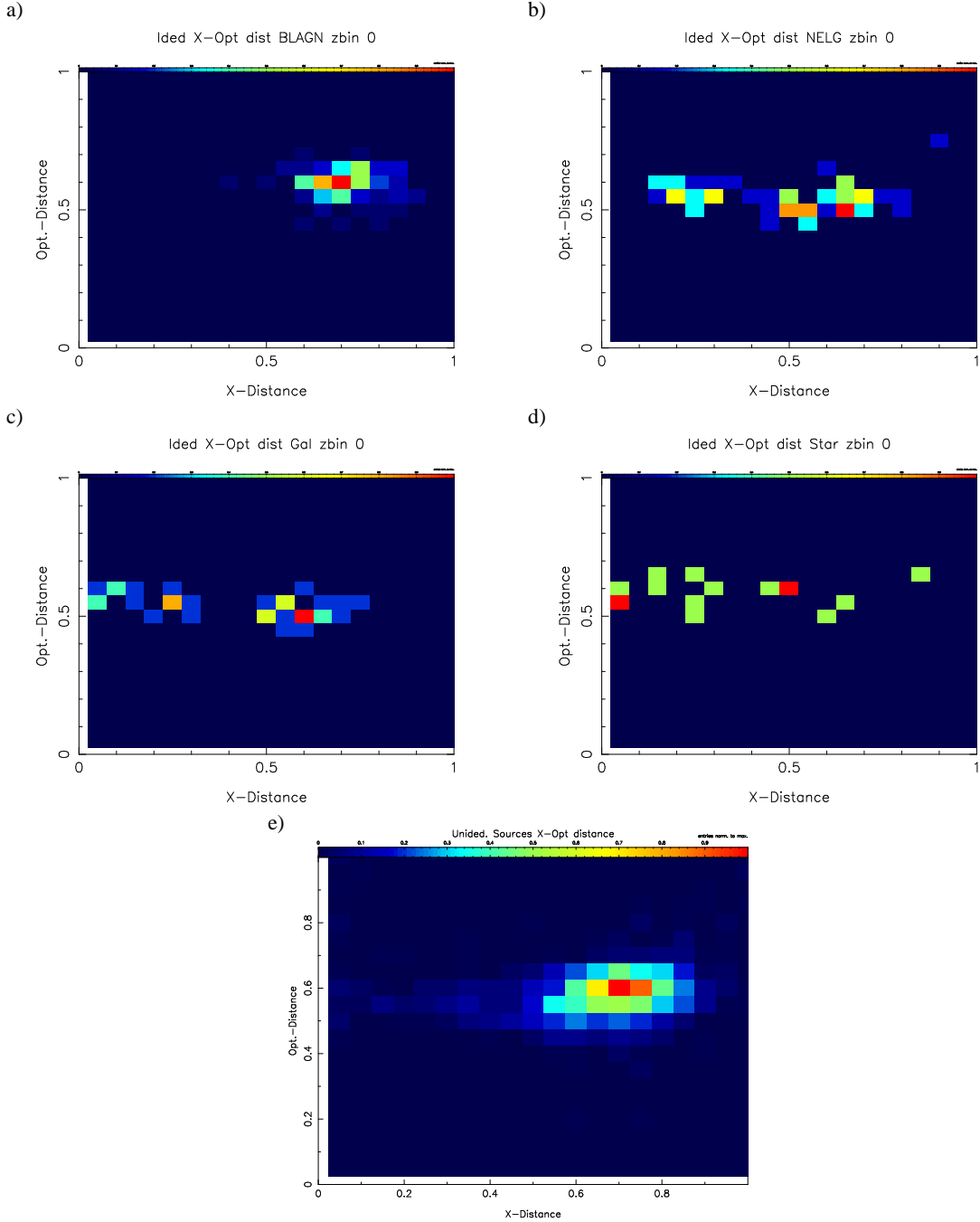

\begin{center}
\begin{tabular}{cc}
a)\includegraphics[height=7cm,angle=-90]{fig3a.eps} & 
b)\includegraphics[height=7cm,angle=-90]{fig3b.eps} \\ 
c)\includegraphics[height=7cm,angle=-90]{fig3c.eps} & 
d)\includegraphics[height=7cm,angle=-90]{fig3d.eps} \\ 
\multicolumn{2}{c}{e)\includegraphics[height=7cm,angle=-90]{fig3e.eps}} 
\end {tabular}
\caption{Color coded distribution of ChaMP sources in the X-ray-Optical distance 
plane. X-ray and optical parameters used 
for calculating distances according to (\ref{distdim}) are respectively 
$\log (f_{x_i}/f_r)$ for $i = S1, S2, H$ X-ray bands, $\log (f_r/10^{-14})$ 
and $\log (f_{opt}/f_r)$ for $opt = z, i, g, u$ bands. Number of sources: 
153 BLAGN, 65 NELG, 37 ALG/Gal and 15 stars. The distribution of 1357 
unclassified ChaMP sources is also shown.}
\label{fig:dist}
\end{center}
\end{figure*}

The result is a distribution in a 2-dimensional space wh-ose 
coordinates correspond to distances in optical and X-ray subspaces as 
defined above. This space is normalized with respect to bin with the 
maximum number of entries and binned. As the parameter space in 
this procedure is truncated, it is more meaningful to normalize it with 
respect to maximum than with respect to total number of entries. This 
normalization somehow amplifies features in the distribution and helps 
classification, whereas normalization to total number of entries makes 
the distribution flatter and featureless. This definition is possible 
because distributions are considered as being the conditional distribution 
in the parameter space. The final distribution is normalized as usual to 
make the total conditional probability equal 1. Distances and 
thereby the distribution of sources in each subspace depends on the 
parameters. Fig. \ref{fig:dist} shows an example of 
distribution of sources.

The conditional probability for an unknown source to belong to a class is 
directly obtained from these distributions. Due to the existence of many empty 
bins in the parameter space, we smear bins by considering not only the 
probability of the bin to which the unknown source belongs, but also its 
neighbors, and determine an average probability. Then, we compare 
this probability with the same quantity for other classes. The source is 
associated to the category with the highest probability. We also request a 
minimum probability of $1/A$ where $A$ is the total surface of the 
2-dimensional distribution\footnote{This corresponds to the probability for 
each bin in a uniform distribution. If objects were distributed uniformly in 
the parameter space, no classification could be done.}. If this condition is 
not satisfied, the source is considered to be unclassifiable.

\begin{figure*}[ht]
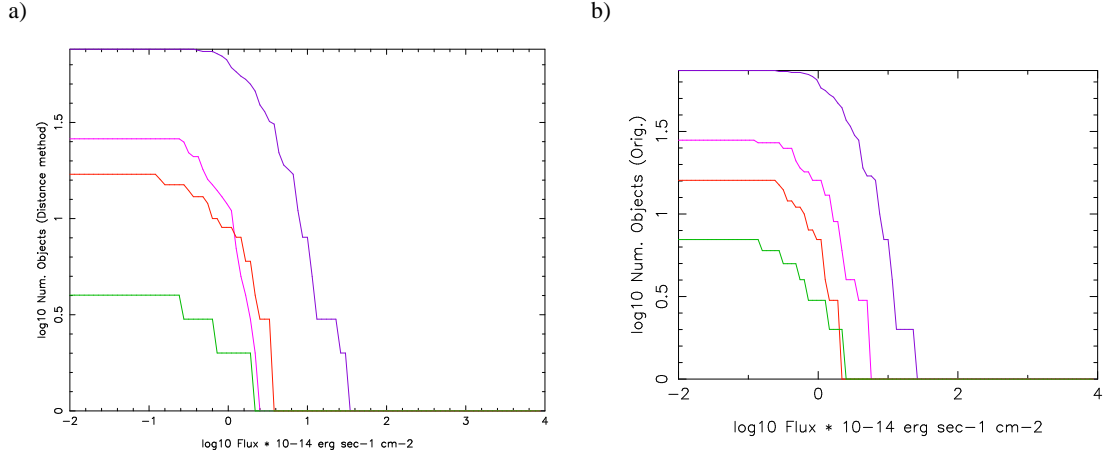

\begin{center}
\begin{tabular}{cc}
a) \includegraphics[height=7cm,angle=-90]{fig4a.eps} & 
b) \includegraphics[height=7cm,angle=-90]{fig4b.eps} 
\end {tabular}
\caption{a) $\log N$ - $\log S$ plot for the objects identified by the 
distance to cluster method, b) the same plot from spectroscopic 
classification.  BLAGN (violet), NELG (magenta), ALG/gal (red), 
stars (green). Parameter space is the same as Fig. \ref{fig:dist}.}
\label{fig:nsdist}
\end{center}
\end{figure*}

The performance of this algorithm is much better than the first one. 
The result is summarized in 
Table \ref{tab:kstest}. Although this method is not able to discriminate 
redshifts, it is fairly good in classification specially when the input set 
is small. Fig. \ref{fig:nsdist} compares the X-ray $\log N$ - $\log S$ 
for 125 classified sources using an input set of 143 sources. The largest 
difference is for stars which are under-represented - only 8 stars in the 
input sample.

\subsection{Low Resolution Spectra} \label{sec:spect}
This method has been inspired from the photometric redshift calculation. 
One way of determining redshift from photometric data is to compare the 
photometric measurements of the source, considered as a low resolution 
spectrum, to templates (\cite{photomz}). Various methods for cla-ssification 
have been used including: analytical parameterization (\cite{zanal}), 
${\chi}^2$-fitting (\cite{photomz}), maximum likelihood (\cite{mlm},\\
\cite{mlm1}), and 
neural network (\cite{zneural}). In this section we discuss ${\chi}^2$-fitting 
and in the next section the maximum likelihood algorithm.

For fitting we have tried two procedures. In the first procedure, we use 
optical/IR and X-ray fluxes of the spectroscopically identified sources to 
determine an average low resolution spectrum, and a dispersion (standard 
deviation) around the average curve for each class and redshift bin 
defined in Table \ref{tab:redshift}. These spectra are used as templates. We 
use the same redshift binning for all the methods. The choice of the bins is 
simply conducted by having a significant number of input sources in 
each bin. The number of bins is also limited by the size of memory necessary 
for buffering the corresponding probability distributions (see 
Sec.\ref{sec:prob}).

The spectrum of each unclassified source is compared to these templates 
according to ${\chi}^2$ difference defined as: 
\bea
\chi^2 = \sum_{i=1}^D \frac {(x_i - X_i)^2}{\sigma_i^2}. \label{chis}
\eea
where $\sigma^2$ includes both the dispersion around the average and 
measurement errors; $X_i$ is the average value of $i^{th}$ parameter in 
the input data set. We note that measurement errors are much smaller than the 
intrinsic dispersion of the population and don't significantly contribute in 
$\sigma^2$. The unclassified source is associated to the class and redshift bin 
with smallest ${\chi}^2$. If all ${\chi}^2 > 1.2 D$ where $D$ is the number 
of freedom degrees, the source is considered to be unclassifiable. 
For $D \leqslant 8$, this is roughly equivalent to $< 30\%$ probability that 
the source belongs to one of the categories but be rejected. 

In the second procedure we simply fit the spectrum of the unclassified sources 
to spectrum of all the members of the available input data set by using 
equation 
(\ref{chis}). For the uncertainty $\sigma$ we have tried measurement errors, 
dispersion as explained for the first procedure, and simple normalization to 
the template. We find that measurement errors are too small as deviation 
estimation, $\chi^2$ becomes too large, and the result of the classification 
becomes biased to the dominant population of sources i.e. to BLAGN (which has 
the smallest dispersion). Two other 
dispersion estimations are better and give more or less the same results. 
In both cases depending on the parameter space - normalisation of fluxes by 
optical or X-ray flux - either stars are over-estimated and ALG underestimated 
or vis-versa. The redshift discrimination quality of these dispersion 
estimation methods are also very similar, although the details of the redshift 
distribution curve can be different. In summary, we can not find any criteria 
preferring one of the dispersion estimation to the other.

The performance of the two spectrum fitting procedures explained here is not 
very different, but fitting to all sources separately is somehow better in 
classifying BLAGNs - the dominant population of the X-ray selected 
sources\footnote{The reason for a better performance of fitting individual 
sources can be the small dispersion of BLAGN spectra. There is more chance to 
find an identified source with very similar spectrum for this class of 
objects.}. Therefore, here we only report the results of this procedure.

\begin{table}[ht]
\caption{Redshift bins\label{tab:redshift}}
\vspace{0.2cm}
\begin{tabular}{lclc}
\hline Band & Redshift & Band & Redshift \\
\hline B1 & z $<$ 10$^{-2}$ & B7 & 1.25 $<$ z $<$ 1.5 \\
B2 & 10$^{-2} <$ z $<$ 0.25 & B8 & 1.5 $<$ z $<$ 1.75 \\
B3 & 0.25 $<$ z $<$ 0.5 & B9 & 1.75 $<$ z $<$ 2 \\
B4 & 0.5 $<$ z $<$ 0.75 & B10 & 2 $<$ z $<$ 2.5 \\
B5 & 0.75 $<$ z $<$ 1 & B11 & 2.5 $<$ z $<$ 3.5 \\
B6 & 1 $<$ z $<$ 1.25 & B12 & z $>$ 3.5 \\
\hline 
\end{tabular}
\end{table}

\begin{figure*}[ht]
\begin{center}
\begin{tabular}{c}
\includegraphics[height=12cm,angle=-90]{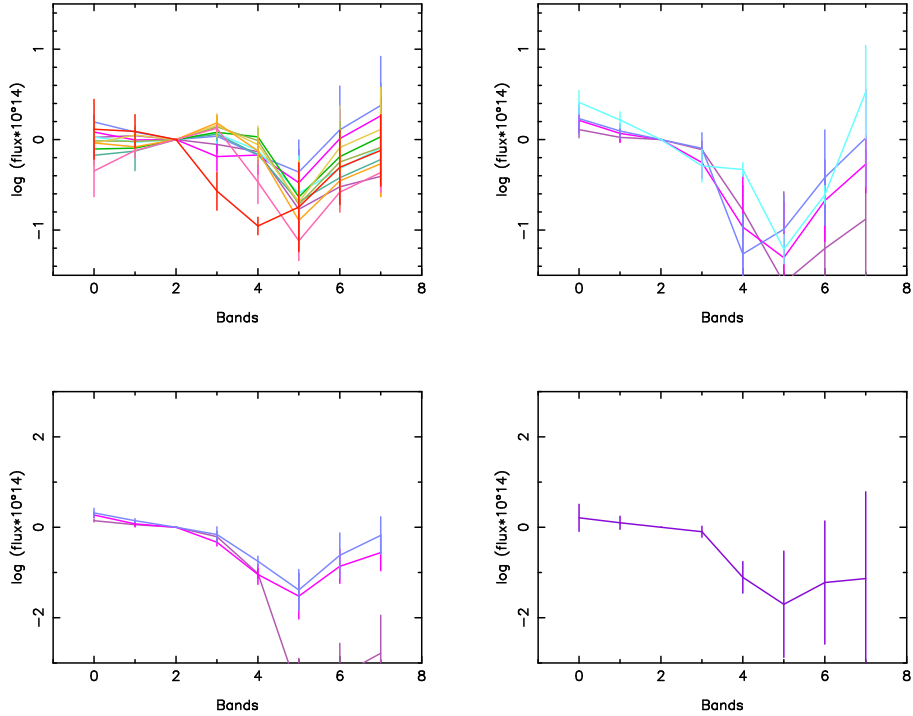} 
\end{tabular}
\caption{Average spectra for each class and redshift. Bars present the 1-sigma 
deviation from average. Parameter space is the same as Fig. \ref{fig:specred}-c. 
Top-left BLAGN, top-right NELG, bottom-left ALG/Gal and bottom-right stars. 
Colors from lowest redshift band to highest are: 
dark violet, violet, blue, cyan, turquoise, green, olive green, gold, orange, 
pink, red, coral.}
\label{fig:specav}
\end{center}
\end{figure*}

The performance of this method is better than distance to clusters. Under 
conditions we have imposed on the fit, the number of unclassified sources 
emerging from this method is only a few percents. The fraction 
of good classifications somehow depends on the quantity used to represent the 
spe-ctrum. We have tried a number of quantities 
including: logarithm of fluxes; relative fluxes to the Chandra broad B-band 
flux $\log (f_j/f_B)$, where $j$ presents all other available X-ray and 
optical/IR bands; relative fluxes to one of the optical fluxes 
$\log (f_j/f_{opt})$; flux ratio of successive bins $\log (f_j /$ $f_{j-1})$ 
(color); and photon number flux (calculated from energy flux) 
$\log (f_j / E_j)$ 
where $E_j$ is the mean energy of the band. We have found that the best 
combinations are relative fluxes to optical $r$ band or to X-ray B band. For 
testing this algorithm we can not use the same data both for input and for 
test. Therefore, all the results in this section are based on completely 
independent input and test sets.

With this algorithm BLAGNs are correctly classified in $> 90\%$ of cases, but 
are under-estimated. There is an over-estimation of NELG where in most cases 
the spectroscopic identification is BLAGN or ALG. Due to under-representa-tion 
of stars in our data set, their classification is in general poor. Depending 
on the parameter space, either they are not correctly classified or there are 
many contaminations from other categories. Nonetheless, our experience with the 
XMM-Newton data shows (\cite{xmmhouri}) that when significant number of stars 
are present in the input data set, this method is well capable of classifying 
them.

It is worth to remind that all the classification methods explained in this 
work are based on conditional (Bayesian) probability. Therefore, the ideal 
case is to have the same number of sources for each class and redshift in the 
input set. In practice however, it is very difficult to collect such a set of 
identified sources as their occurrence in the serendipitous data is quite 
different.

\begin{figure*}[ht]
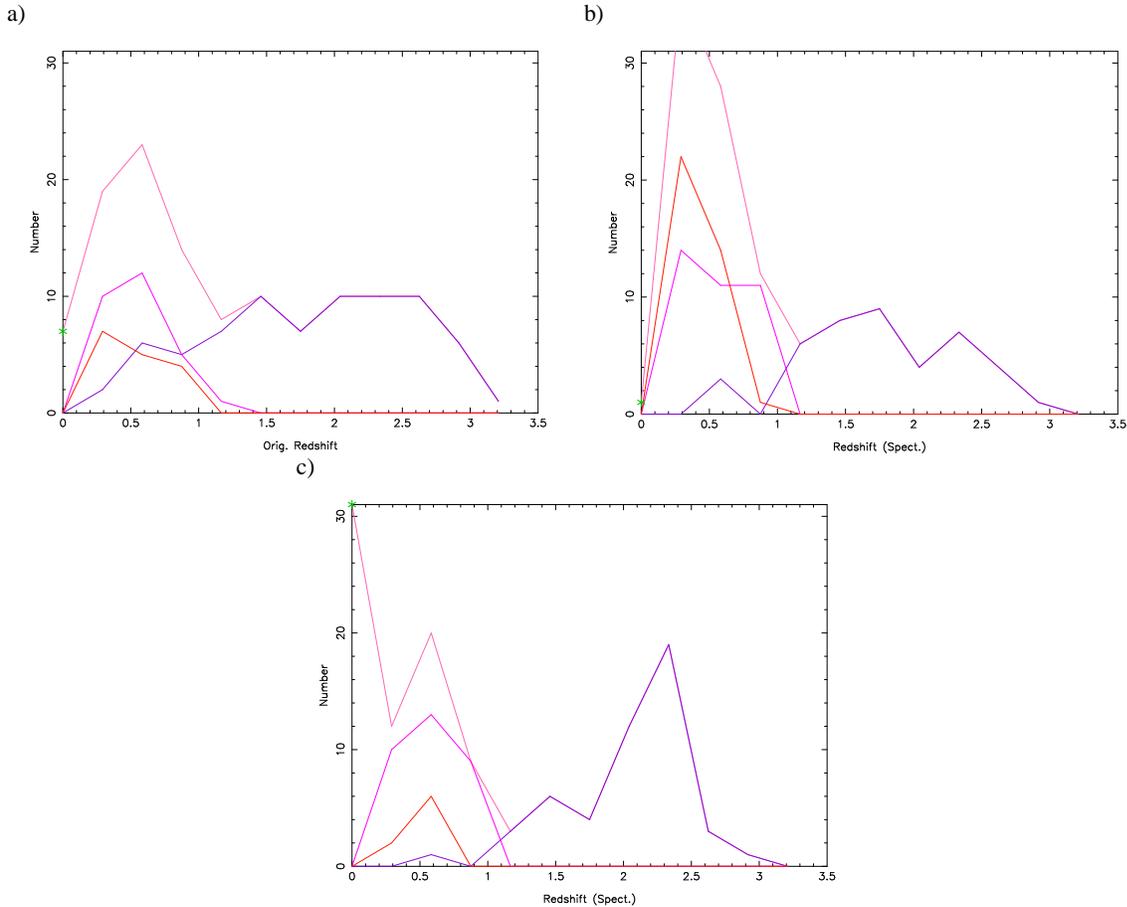

\begin{center}
\begin{tabular}{cc}
a)\includegraphics[height=7cm,angle=-90]{fig6a.eps} & 
b)\includegraphics[height=7cm,angle=-90]{fig6b.eps} \\ 
\multicolumn{2}{c}{c)\includegraphics[height=7cm,angle=-90]{fig6c.eps}} 
\end {tabular}
\caption{Redshift distribution of sources: a) Spectroscopic and from 
automatic classification by low resolution spectrum method with ratio of 
fluxes to X-ray B band b) and to optical $r$ band c). BLAGN (violet), 
NELG (magenta), ALG/gal (red) and stars (green). Stars at redshift zero are 
shown by $\times$. Average deviation (see explanation in the text) is used in 
the fit (see Table \ref{tab:kstest} for details). Note that depending on the 
parameters, in one case stars are over-detected and in the other case 
under-detected. We also note the difference between features in redshift 
distribution. Simple statistical tests such as KS or $\chi^2$ do not 
show these detail differences.}\label{fig:specred}
\end{center}
\end{figure*}

The precision of the recovered redshift in this method is more modest 
$\sim 40\%$. Nonetheless, $\sim 25\%$ of incorrect redshifts have only 
$\pm 1$ bin of difference from the spectroscopic redshift which, 
for $z < 2$, means $\pm 0.25$. Fig. \ref{fig:specred} compares the 
redshift distributions of classified sources by this method with the 
spectroscopic ones. To see if reducing the number of redshift bins improves the 
performance of redshift determination, we have also tested this algorithm with 
only 6 redshift bins with $\Delta z = 0.5$. We find that the performance is 
practically the same (see Table \ref{tab:kstest}). 
The reason can be the fact that by averaging/binning spectra in too large 
intervals we smear critical redshift dependent features.

Because this method is used by many optical/IR surveys (\cite{photomz}) for 
photometric redshift determination, it is worth to mention that the reason for 
their performance is the large input sets of thousands or even tens of thousand 
of spectroscopically classified spectra. At present such an input set is not 
available for X-ray surveys, and therefore they should use more ingenious 
algorithms for classification.

\subsection{Parameters as Independent Measurements} \label{sec:ml}
\begin{figure*}[ht]
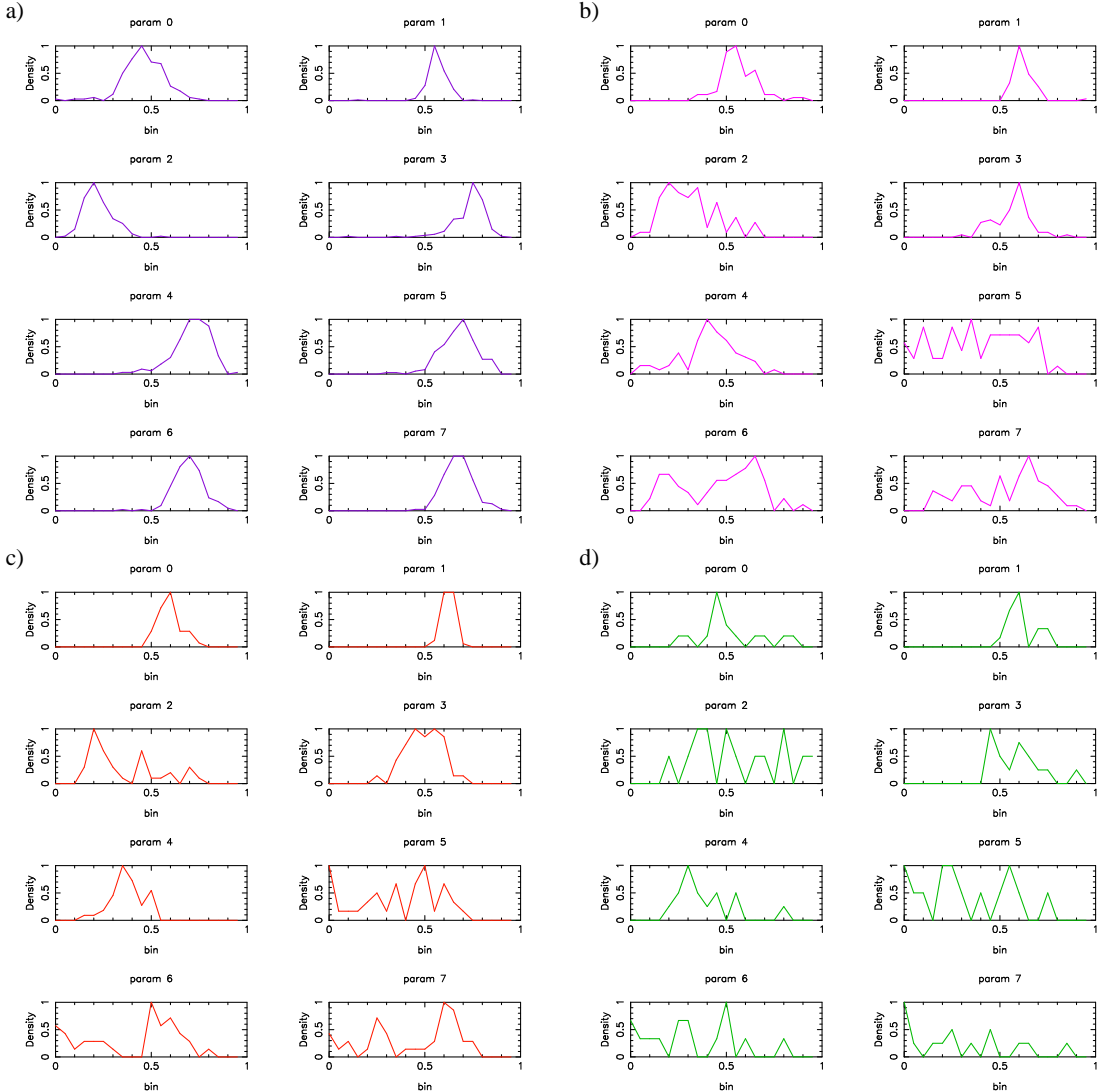

\begin{center}
\begin{tabular}{cc}
a)\includegraphics[width=7cm,angle=-90]{fig7a.eps} & 
\hspace{-1.5cm}b)\includegraphics[width=7cm,angle=-90]{fig7b.eps} \\
c)\includegraphics[width=7cm,angle=-90]{fig7c.eps} & 
\hspace{-1.5cm}d)\includegraphics[width=7cm,angle=-90]{fig7d.eps} 
\end {tabular}
\caption{Distribution of parameters (no redshift discrimination) for ChaMP 
classified sources: BLAGN (violet), NELG (magenta), ALG/gal (red) and 
stars (green). Parameters are the same as in Fig.\ref{fig:dist}.} 
\label {fig:mldistr}
\end{center}
\end{figure*}

This popular method for hypothesis/model testing uses each measurement as an 
independent observation of the hypothesis/model under scrutiny. For 
classification of astronomical sources, the measured value of various physical 
parameters can play the role of independent observations~\footnote{The way we 
use maximum likelihood here is somewhat different to its usual application. 
For parameter estimation, usually one quantity is measured for fixed values of 
another parameter. The distribution for all the measurements is assumed to be 
the same. A good example is the application of ML to Cosmic Microwave 
Background (CMB) anisotropy for estimation of cosmological parameters. In this 
case the observed quantity is the anisotropy at different angular scales. In 
classification problem we have one measurement but multiple quantities with 
different distributions. Therefore the mathematical meaning of the likelihood 
here is not exactly the same as usual. This is not a concern because our aim 
is classification of objects and not their statistics. Therefore, any means 
that can discriminate them is appreciated.}. We should not however forget that 
usually these parameters - in our case X-ray and optical fluxes - are not 
independent. Nonetheless, their relations are subtle and linear methods 
such as principal component analysis are not appropriate for finding 
independent combinations. Moreover, different classes can have different 
sets of independent 
parameters. This makes the definition of one set of parameter for all classes 
impossible. Therefore, we simply use this method without any attempt to reduce 
the number of parameters or find an independent set. We neglect their 
correlations and use the maximum likelihood method to find the probability 
distribution for each parameter, each class and each redshift. An example of 
these 1-dimensional distributions is shown in Fig. \ref{fig:mldistr}. The main 
conclusion we make from this figure is that NELGs, ALGs and stars have very 
similar distributions. BLAGNs are somehow different from other 
categories, but they cover the same region of the parameter space, and there 
is no special feature capable of singling out 
just one category of sources. This conclusion is also consistent with 
2-dimensional distributions in Figs. \ref{fig:hf}. 

For each unclassified source we calculated its likelihood to belong to a class 
and redshift bin as the following:
\bea
{\mathcal L} = \sum_{i=1}^D \log (f_i (x_{i})). \label {ml}
\eea
where $f_i$ is the distribution of $i^{th}$ parameter calculated using 
spectroscopically identified sources and $x_i$ is the measured value of 
$i^{th}$ parameter for the unclassified source. Similar to the distance method 
we request a minimum value for likelihood ${\mathcal L}$. Sources with 
likelihoods less than this lower limit for all categories and redshift bins 
are considered as unclassifiable.

Mathematically, ML method and $\chi^2$-fitting explained in the 
Sec.\ref{sec:spect} are both based on 1-dimensional distribution of 
parameters. Their 
difference when they are applied to classification problem is in the way 
they treat the input data. In $\chi^2$-fitting if we fit the spectrum of an 
unknown source to a template for each class, it is equivalent to 
considering a 1-dimensional Gaussian distribution for each parameter around 
the corresponding value for the template. From Fig. \ref{fig:mldistr} it is 
clear that parameters used here don't have Gaussian distributions. If we fit 
the spectrum of an unclassified source to all the available sources to find the 
most similar source, one of them eventually will have the smallest $\chi^2$, 
but we don't explore the statistical properties of the available input set 
i.e. how frequently one type of spectrum occurs. Moreover there is no way to 
fill the gap between different templates by for instance interpolation.
In contrast, ML method is not based on any predefined distribution for 
parameters and gets the information directly from the learning sample. 
As a result the performance of this method is much better than 
$\chi^2$-fitting.

We use the same parameter space as that adopted for the low resolution 
spectrum method, i.e. either $\log (f_j/f_B)$ or $\log (f_j/f_r)$ with $j$ 
indicating other X-ray or optical/IR bands. For binning each measurement, 
the range of possible values must be defined. We fix the range for each 
parameter such that it is covered by the largest input data set available. 
Evidently, for unclassified sources we can not guarantee that they will fall in 
the same range. Therefore apriori we should consider a larger range. But the 
input probability distribution in these regions would be zero and will not 
help the identification. In practice when a measurement is out of range, 
we associate it to the closest bin i.e. to the first or the last bin. In any 
case the definition of the volume of the parameter space affects 
classification and we have not found an optimum prescription. This issue is 
relevant for both 1-dimensional and multi-dimensional distributions.

\begin{figure*}[ht]
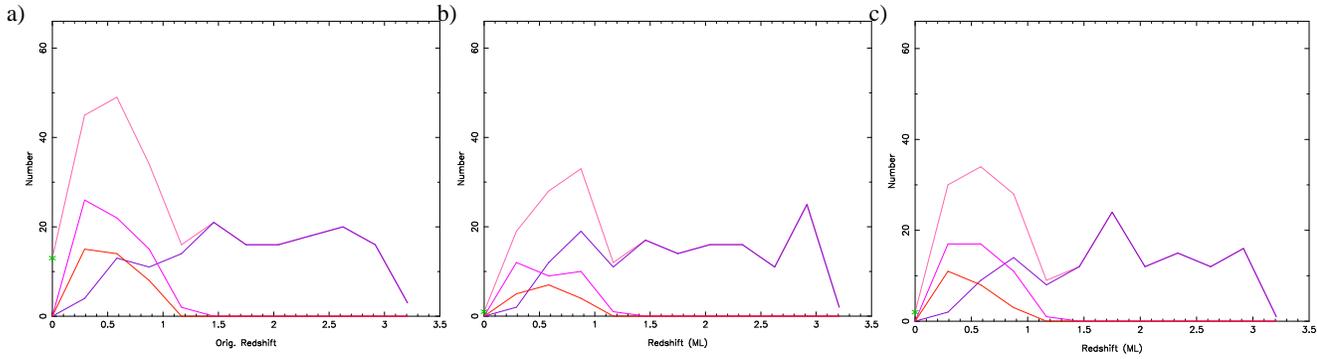

\begin{center}
\begin{tabular}{ccc}
\hspace{-0.7cm}a)\includegraphics[height=5.7cm,angle=-90]{fig8a.eps} 
\hspace{-0.3cm}b)\includegraphics[height=5.7cm,angle=-90]{fig8b.eps} 
\hspace{-0.3cm}c)\includegraphics[height=5.7cm,angle=-90]{fig8c.eps} 
\end {tabular}
\caption{Redshift distribution of sources: a) Spectroscopic, b) 
Automatic classification by maximum likelihood method with ratio of fluxes to 
X-ray $B$ band, c) The same as b) but with flux ratio to optical $r$ band. 
Definition of curves are the same as Fig. \ref{fig:specred}. See Table 
\ref{tab:kstest} for details.} 
\label{fig:mlz}
\end{center}
\end{figure*}

The performance of the maximum likelihood method is very good, for both a 
low number 
of unclassified sources and good estimation of redshift distribution, see 
Table \ref{tab:kstest} and Fig. \ref{fig:mlz}. Like previous methods, the 
performance for classification of stars is poor, but we attribute this to the 
under-representation of the stars in our input set rather than the fault of the 
algorithm. 

Because this method is based on 1-dimensional distributions, we can use the same 
data set for input and for test. This is equivalent to assuming distributions 
to be exact, i.e. when the size of the input set goes to infinity. Tests with 
independent sets also show roughly the same level of performance.

\begin{figure*}[ht]
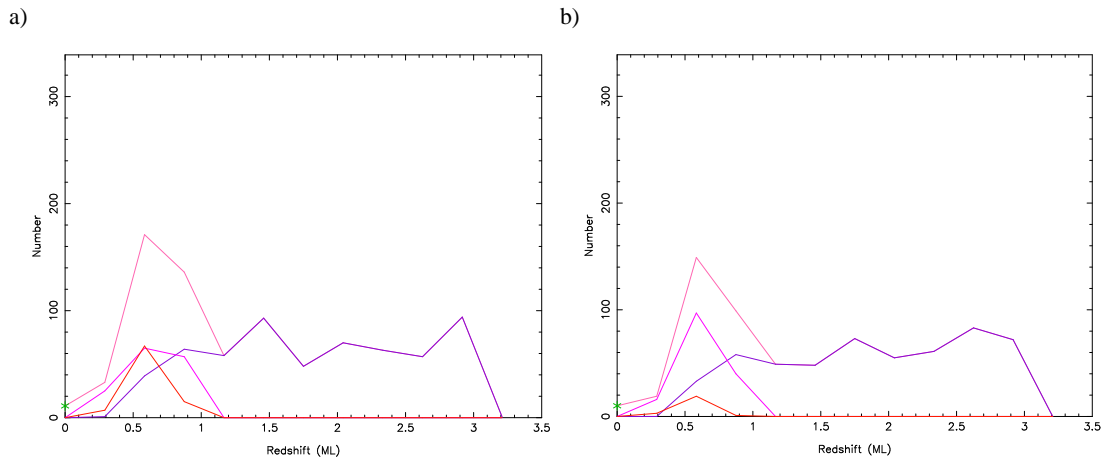

\begin{center}
\begin{tabular}{cc}
a)\includegraphics[height=7cm,angle=-90]{fig9a.eps} 
b)\includegraphics[height=7cm,angle=-90]{fig9b.eps} 
\end {tabular}
\caption{Redshift distribution of 1357 ChaMP sources classified by maximum 
likelihood method with ratio of fluxes to X-ray $B$ band a) and with optical 
$r$ band b). Input set is the same as Fig. \ref{fig:mlz}-a. Definition of 
curves is the same as Fig. \ref{fig:specred}. }\label{fig:unidmlz}
\end{center}
\end{figure*}

Despite the efficiency of this method for moderate size input set, it is not 
sure that it keeps the same performance for large number of sources when 
1-dimensional distributions approaches uniformity. The evidence for such 
concern comes from the classification of 1357 unknown ChaMP so-urces in 
Fig \ref{fig:unidmlz} by this method. Redshift distribution of 718 sources 
that got classification looks very similar to the input set. If our input 
distribution was very close to the cosmological distribution, we evidently 
expected such behaviour. But we know that our input data set of 268 sources 
is very far from being the unbiased cosmological distribution, and therefore 
the similarity of the redshift distribution to the input is an evidence of 
fast saturation of the algorithm. The problem of having close to uniform 
1-dimensional distributions can be solved by making finer redshift bins. 
Apriori the bias toward the input distribution is not a problem if the 
input set is unbiased. Otherwise, we should expect a relative deterioration of 
classification for large input data sets. This seems somehow counterintuitive 
because classification of each source is independent of others. The 
deterioration is however due to the accumulation of errors, otherwise the 
probability of wrong classification per source is always the same.

\subsection{Multi-dimensional Probability} \label{sec:prob}
The space of all independently measured quantities contains all the 
available information about sources and includes all the correlations between 
parameters which can be used for discriminating between various 
classes (\cite{bayes}). As we mentioned before, although based on physical 
argument-s there must be a relation between various fluxes, their correlations 
are complicated and nonlinear. The visual demonstration of the complexity of 
distribution in a multidimensional parameter space is difficult. Nonetheless, 
the diversity of low resolution spectra (see Fig. \ref{fig:specav}) and 
2-dimensional projections such as Fig. \ref{fig:dist} and Fig. \ref{fig:hf} 
are the evidence for 
this claim. A linear relation between a subset of $d$ parameters in a 
$D-$dimensional parameter space makes sources cluster in a $D - d$ 
dimensional subspace. If data has such a structure, it must show up at least 
in some of the 2-dimensional projections, and the distribution of sources 
should have an approximately axial symmetry in the direction of the projected 
principal component. We have tried various combination of 2-dimensional 
projections and find no symmetries permitting a reduction of the parameter 
space. This means that application of a dimensional reduction algorithm like 
principal component does not help to reduce the volume without losing 
important information. The only possible optimization of the volume of the 
parameter space is therefore based on the choice of the range of parameters, 
as mentioned in the previous section.

For exploring the multi-dimensional distribution of sou-rces in the parameter 
space, we are obliged to restrict its dimension and/or the number of bins for 
each parameter\footnote{In the current versions of the GNU $C^{++}$ compiler, 
the memory allocation is limited to 2 GB corresponding to 4-byte addressing, 
and the total allocated buffer by the code can not exceed this value.}. We 
have found that the performance is much more sensitive to X-ray than optical 
flux binning. The reason seems to be a higher variation of X-ray fluxes than 
optical ones (see Fig. \ref{fig:specav}). The reason can be simply the wider 
band-width 
of X-ray - about 1.5 orders of magnitude - than optical filters band-width 
which, from $z$ to $u$, is less than 0.5 orders of magnitude. When 4 or 5 
optical fluxes are used, we consider 6 bins for each optical flux ratio and 8 
bins for each X-ray one. For less than 4 optical fluxes, 8 bins are used for 
optical and 10 for X-ray fluxes. 
\begin{figure*}[ht]
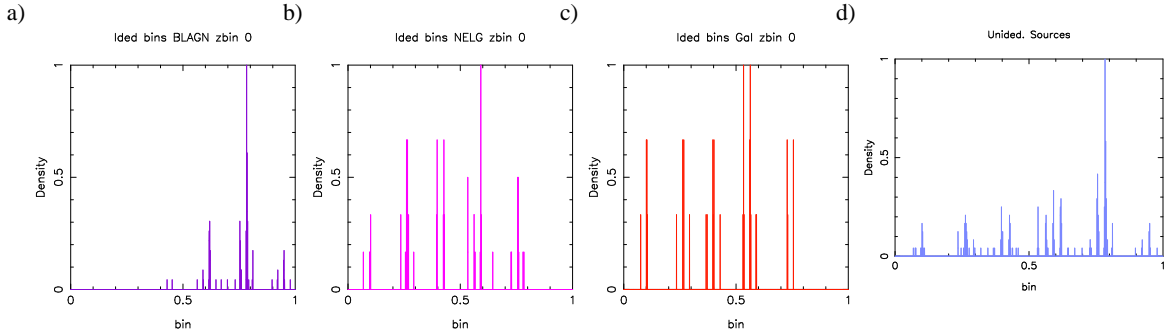

\begin{center}
\begin{tabular}{cccc}
\hspace{-1.cm}a)\includegraphics[height=5cm,angle=-90]{fig10a.eps} & 
\hspace{-2cm}b)\includegraphics[height=5cm,angle=-90]{fig10b.eps} & 
\hspace{-2cm}c)\includegraphics[height=5cm,angle=-90]{fig10c.eps} & 
\hspace{-2cm}d)\includegraphics[height=5.5cm,angle=-90]{fig10d.eps} 
\end{tabular}
\caption{Vector presentation of the multi-dimensional parameter space (all 
redshifts): a) BLAGN, b) NELG, c) ALG/gal, d) 1357 unclassified ChaMP sources. 
It is obtained by sweeping bins in each axis. Semi-periodic structure of the 
plot reflects the similarity between the content of neighbour bins. This 
illustration helps to see the difference between multi-dimensional distribution 
of astronomical classes. For the sake of presentation these plots are made 
with only 4 parameters and no redshift discrimination.} \label{fig:bins}
\end{center}
\end{figure*}

After distributing spectroscopically identified sources in each class and 
redshift bin, distributions are normalized in the same way explained in 
Sec.\ref{sec:proj} and considered to 
be the multi-dimensional probability density for each class and redshift. 
To classify an unclassified source, its place in the parameter space is 
determined in the same way and its probability to belong to a class and a 
redshift bin is determined from multi-dimensional distributions. Due to the 
small nu-mber of sources in the input, we use a smeared probability i.e. for 
each class we don't use only the probability of the bin to which the 
unclassified source belongs but also the closest neighbour bins and determine 
an average probability (without weighting). Averaging can be removed if the  
input set is sufficiently large. We also consider a minimum probability 
inversely 
proportional to the total volume of the parameter space. If for all classes 
and redshifts the probability is smaller than this limit, we consider the 
source as unclassifiable.

\begin{figure*}[ht]
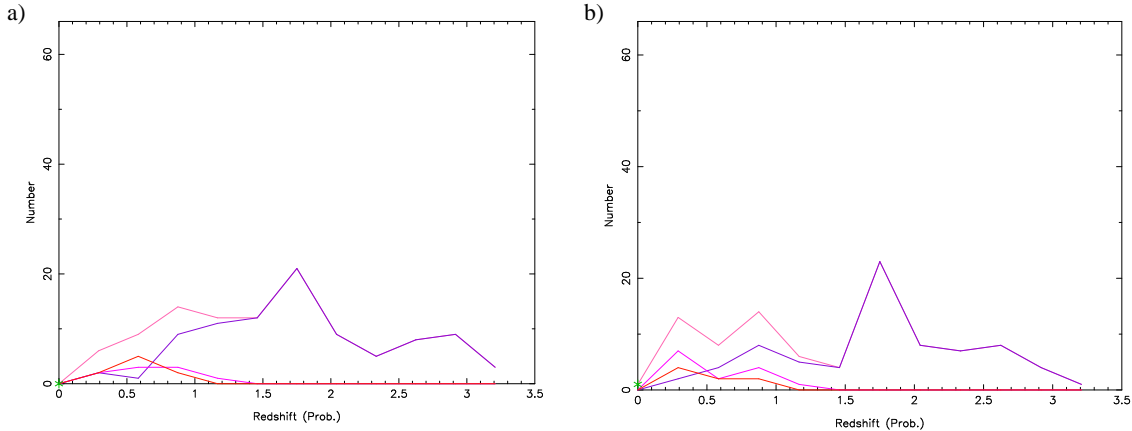

\begin{center}
\begin{tabular}{cc}
a)\includegraphics[height=7cm,angle=-90]{fig11a.eps} & 
b)\includegraphics[height=7cm,angle=-90]{fig11b.eps} 
\end{tabular}
\caption{Redshift distribution of sources automatically classified by 
multi-dimensional probability method a) with ratio of fluxes to X-ray $B$ band 
and b) with optical $r$ band. The spectroscopically determined distribution 
is the same as Fig. \ref{fig:mlz}-a. Definition of curves is the same as 
Fig. \ref{fig:specred}. See Table \ref{tab:kstest} for details.} 
\label{fig:probz}
\end{center}
\end{figure*}

Despite the large parameter volume of this method and relatively small number 
of input sources available, it has the best performance both in classification 
of sources and in determination of redshifts. When the same data set is used 
for input and test - equivalent of having either the exact distribution or 
when the size of input set is very large - the identification performance is 
close to $\sim 98\%$ and redshift classification $\sim 80\%$ (see Table 
\ref{tab:kstest} for details). When independent input and test sets are used, 
many sources are left unclassified and the performance of redshift 
classification is more modest. This has a serious impact on the redshift 
distribution curve, see Fig. \ref{fig:probz}. One can therefore conclude that 
for this method a much larger input data set is necessary. Evidently, this is 
not a surprise as the volume of the parameter space is much larger than when 
1-dimensional distributions are used.

\begin{figure*}[ht]
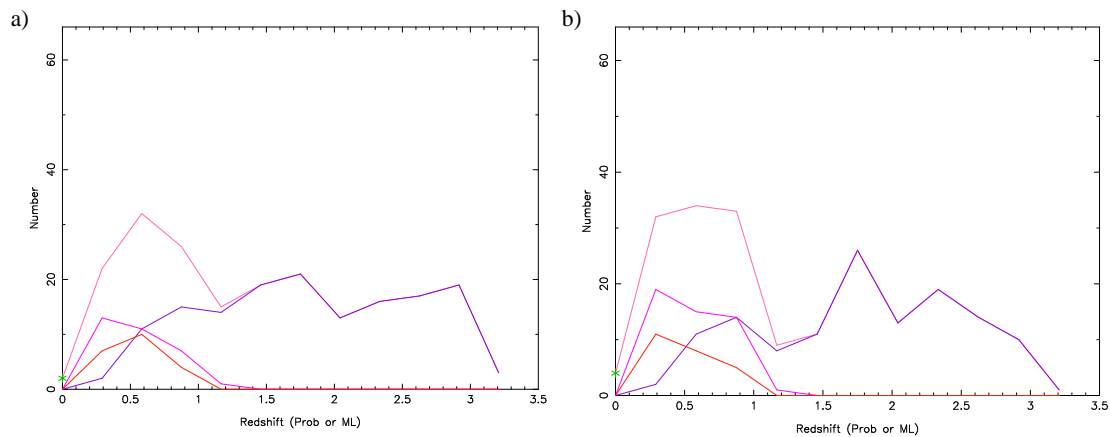

\begin{center}
\begin{tabular}{cc}
\hspace{1cm}
a)\includegraphics[height=7cm,angle=-90]{fig12a.eps} 
b)\includegraphics[height=7cm,angle=-90]{fig12b.eps} 
\end {tabular}
\caption{The same as Fig. \ref{fig:probz}, but when a source is unclassifiable 
by multi-dim distributions, classification by maximum likelihood or low 
resolution spectrum is used. Note that due to under-representation of stars, 
the improvement in their detection is not significant. Parameter space of 
a) and b) are respectively the same as a) and b) in Fig. \ref{fig:probz}} 
\label{fig:probmlz}
\end{center}
\end{figure*}

The problem of unclassified sources can be partially sol-ved if we use 
multi-dimensional probability along with ma-ximum likelihood and low 
resolution spectrum. Fig. \ref{fig:probmlz} sho-ws the redshift distribution 
of statistically classified sources by applying first multi-dimensional 
probability and when a source can not be identified by this method applying, 
by order of priority, ML or low resolution spectrum fit. In this way, not only 
many of unclassified sources get classified, but also the result of the KS and 
$\chi^2$ tests show that the redshift distribution becomes much closer to the 
one from spectroscopy. Fig. \ref{fig:probmlz} shows the distribution of the 
same sources in the X-ray-optical flux plane and demonstrates how the 
complementary use of multiple methods can improve classification. Nonetheless, 
our tests with various divisions of the available data set and choices of 
parameters show that when multiple methods with various degree of precision 
are used in a complementary manner, the contribution of less precise 
algorithms must be kept small otherwise they reduce the overall performance. 
They should be applied only when more precise methods can not classify a 
source, and if possible, other constraints such as higher minimum probability 
should be applied.

We should also mention that the 
large amount of memory necessary for buffering each distribution is an 
important limit for application of this method to finer redshift division even 
when in each redshift bin enough input data is available. The solution can 
be a multi-step classification: First sources are classified in a category and 
relatively large redshift band. Then, multi-dimensional probability or another 
method such as maximum likelihood can be used to classify them in finer 
redshift bands. 

\begin{figure*}[ht]
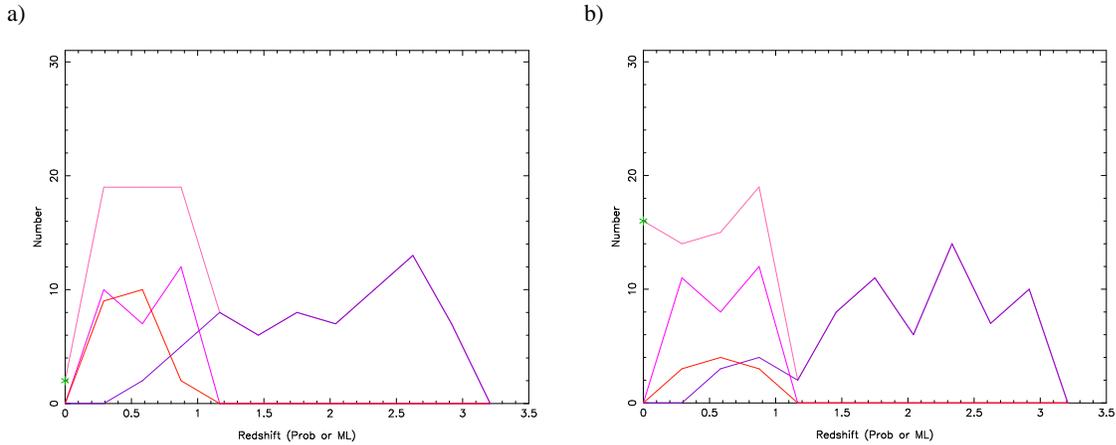

\begin{center}
\begin{tabular}{cc}
a)\includegraphics[height=7cm,angle=-90]{fig13a.eps} & 
b)\includegraphics[height=7cm,angle=-90]{fig13b.eps} 
\end {tabular}
\caption{Redshift distribution of automatically classified sources by 
complementary use of 3 methods as explain in the text, a) with ratio of 
optical 
$g$ and $r$ fluxes and 3 X-ray bands to X-ray $B$ flux and b) with ratio of 
optical $g$, $r$ and $i$ bands to $r$ band flux. The spectroscopically 
determined distribution is the same as Fig. \ref{fig:mlz}-a. Definition of 
curves is the same as Fig. \ref{fig:specred}. See Table \ref{tab:kstest} for 
details.} \label{fig:nou}
\end{center}
\end{figure*}

Finally, we discuss the role of various optical filters in the statistical 
classification of X-ray selected sources. This issue is important for 
optimizing the time and effort needed for optical follow-up and imaging. 
Fig. \ref{fig:nou} shows the redshift distribution of sources when only  
$g$, $r$ and $i$ optical fluxes are used. We find that the impact 
on the classification is not very significant, but the redshift distribution 
is somehow affected. 

\subsection {General Notes about Statistical Classification} 
\label {sec:gennote}
A general characteristic of all the methods we have discussed in this work 
is that although they are based on learning - no predefined rule or relation 
between parameters of a class is implemented in the algorithms - they don't 
extrapolate the input knowledge. If a source has parameters in a part of the 
parameter space uncovered by the input data, there is no way to guess its 
class. Averaging the probability in neighbour bins can partially compensate 
for the lack of information, but it is not useful when the input distributions 
are very disconnected because of their limited size. The same type of problem 
exists for nearest neighbour algorithm in which although isolated entries 
can be always related to a nearest class, the reality of the association is 
very doubtful.

One way of improving the knowledge about the distribution of classes in the 
parameter space is to add statistically classified sources to the initial 
input data and re-access the probability distribution. When the distribution 
is not too disconnected this procedure helps to extend the knowledge to part 
of the parameter space uncovered by the initial data and to improve the 
probability estimation. The weak point of this procedure is that if initially 
the classification performance is uncertain, by adding incorrectly classified 
sources to the learning set, errors accumulate and deteriorate the 
classification performance.

\begin{figure*}[ht]
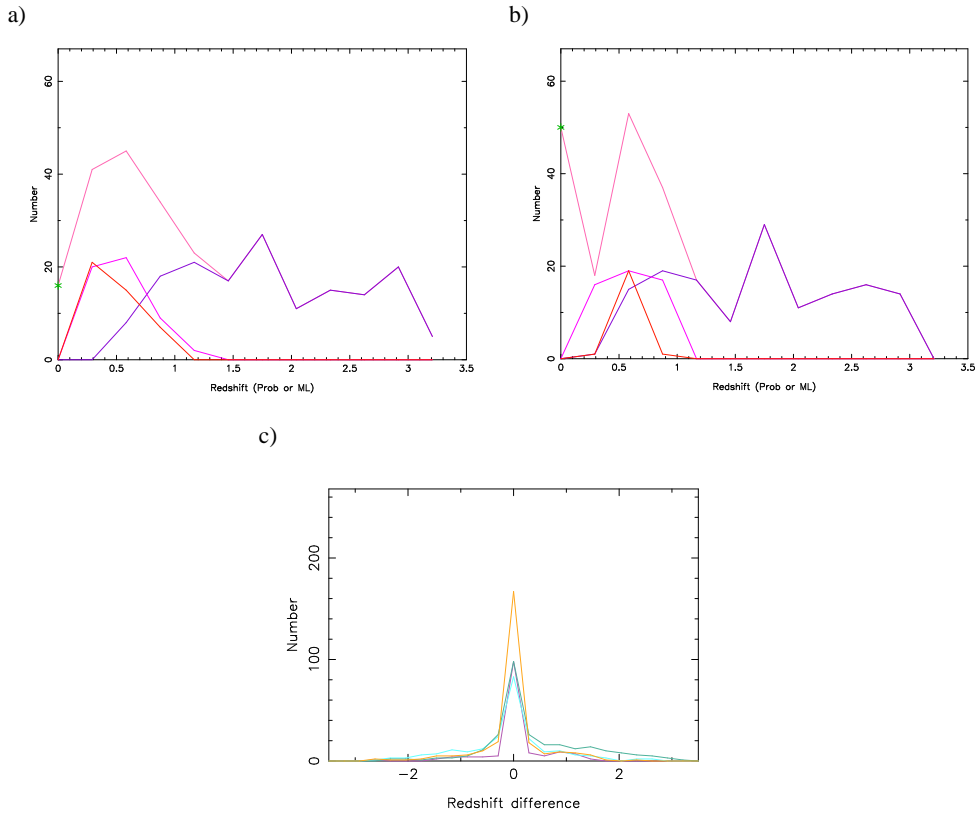

\begin{center}
\begin{tabular}{cc}
a)\includegraphics[height=6cm,angle=-90]{fig14a.eps} & 
b)\includegraphics[height=6cm,angle=-90]{fig14b.eps} \\ 
\\
\multicolumn{2}{c}{c)\includegraphics[height=6cm,angle=-90]{fig14c.eps}} 
\end {tabular}
\caption{Reclassification of 268 spectroscopically identified ChaMP sources.
The input data comes from statistically classified sources using the three 
algorithm as explained in the text. Parameter space for obtaining a) and b) are 
respectively the same as a) and b) in Fig.\ref{fig:probz} and the same are used 
for reclassification. c) Difference between spectroscopic and statistical 
redshift estimation for a), low resolution spectrum (dark green), 
maximum likelihood (cyan), multi-dimensional probability (violet), 
complementary use of three methods (orange).} 
\label{fig:unidflux}
\end{center}
\end{figure*}

To test the possibility of knowledge extension and the performance of 
statistical classifications of unknown Cha-MP sources shown in 
Fig. \ref{fig:unidmlz}, we use these sources as input data to the 
classification 
algorithm and statistically classify the spectroscopically identified sources. 
We make two separate data sets for the parameter spaces described in 
Fig. \ref {fig:probz}. We note that according to this plot and 
Table \ref{tab:kstest} when the ratio of fluxes to optical $r$ flux are used, 
there is a clear bias toward stars and under-representation of ALG/gal in the 
classification. Consequently, we expect that the corresponding statistically 
classified sources as input to the algorithms must be a poorer classifier than 
when the parameter space is define by ratio of flux to the X-ray B band. 
This is exactly what we find when we use these sets to reclassify 
spectroscopically identified sources, see Fig. \ref{fig:unidflux}. 
For less biased parameter space of flux ratios to X-ray B band as 
input set, the number of detected stars is very close to the exact value (16 in 
place of 15), in contrast, to the cases when all or part of the ChaMP 
identified sources have been used as input. The parameter space defined from 
ratio of flux to optical $r$ band over estimates stars.

As we mentioned before, stars are under-represented in our set of 
spectroscopically identified sources. In the statistically classified set their 
fraction is more significant (see Table \ref{tab:kstest} for details). This 
leads to a better classification of stars when this set is used for learning. 
This is clear evidence of {\it knowledge extension}. Moreover, this 
investigation hints to a methodology for testing the quality of classification 
of unknown objects by inverting the place of learning and output data sets.

\begin{figure*}[h]
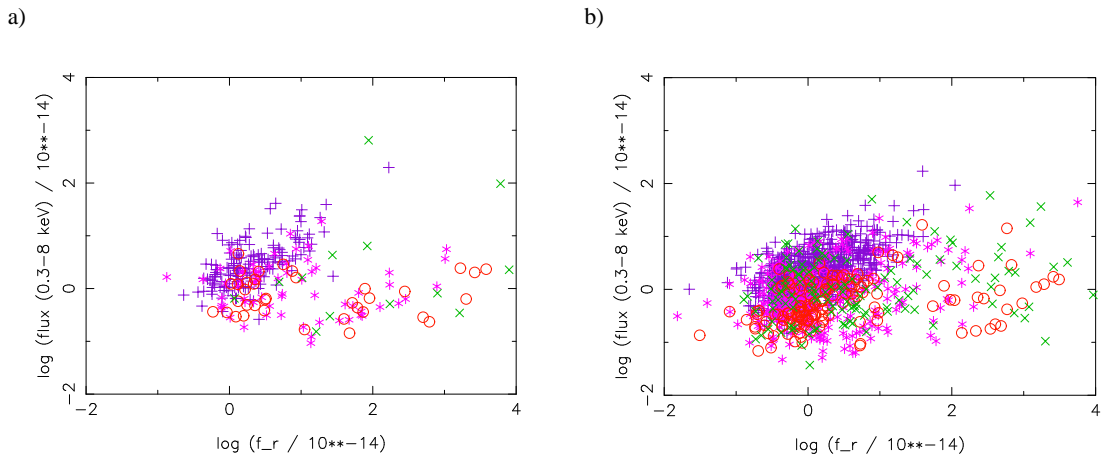

\begin{center}
\begin{tabular}{cc}
a)\includegraphics[height=7cm,angle=-90]{fig15a.eps} & 
b)\includegraphics[height=7cm,angle=-90]{fig15b.eps} 
\end {tabular}
\caption{Distribution of ChaMP identified sources a) spectroscopically and b) 
statistically, in optical r and X-ray B band plane. Definition of colors and 
symbols are the same as Fig. \ref{fig:hf}.}\label{fig:xxfxidfopt}
\end{center}
\end{figure*}

Another strategy for knowledge enhancement that can be also applied to 
disconnected distributions is automatic rule detection. This concept is in 
some ways similar to automatic cluster detection. However, attempts here should 
be concentrated on finding extendable features like symmetries because 
clustering is already included in the probability distribution. For instance, 
searching the principal component for a subclass, a class or multiple classes 
can help to find a symmetry axis for that group of sources. 
Fig. \ref{fig:xxfxidfopt} shows that BLAGN have a roughly oblique 
elliptical distribution in X-ray-optical flux plane. Assuming that this 
behaviour is extendable to a larger sample, one can use the closeness to 
symmetry (principal component) axis to classify sources. We leave this 
investigation to a future work when more 
spectroscopically identified sources are available. Note also that this 
figure confirms that we can not use principal component analysis to reduce the 
number of parameters or find uncorrelated quantities for all the X-ray sources 
because different classes have different set of approximately independent 
parameters.

\section{Effect of Statistical Classification on Physical Conclusions}
Irrespective of the method used for statistical classification, the input set 
leaves its fingerprint on the physical conclusions. Therefore, only if the 
input data is sufficiently - with but a few percent of fluctuations - 
representative of the wh-ole population, one can be confident on the 
reliability of the deduced conclusions about the physical behaviours from 
statistical classification. By giving some examples in this section, we show 
that the available data set is yet too small and a much larger input set is 
needed. At the same time, these examples clarify how one should investigate 
artifacts left in the output of the statistical classification from the 
input data or classification methods.

\begin{figure*}[p]
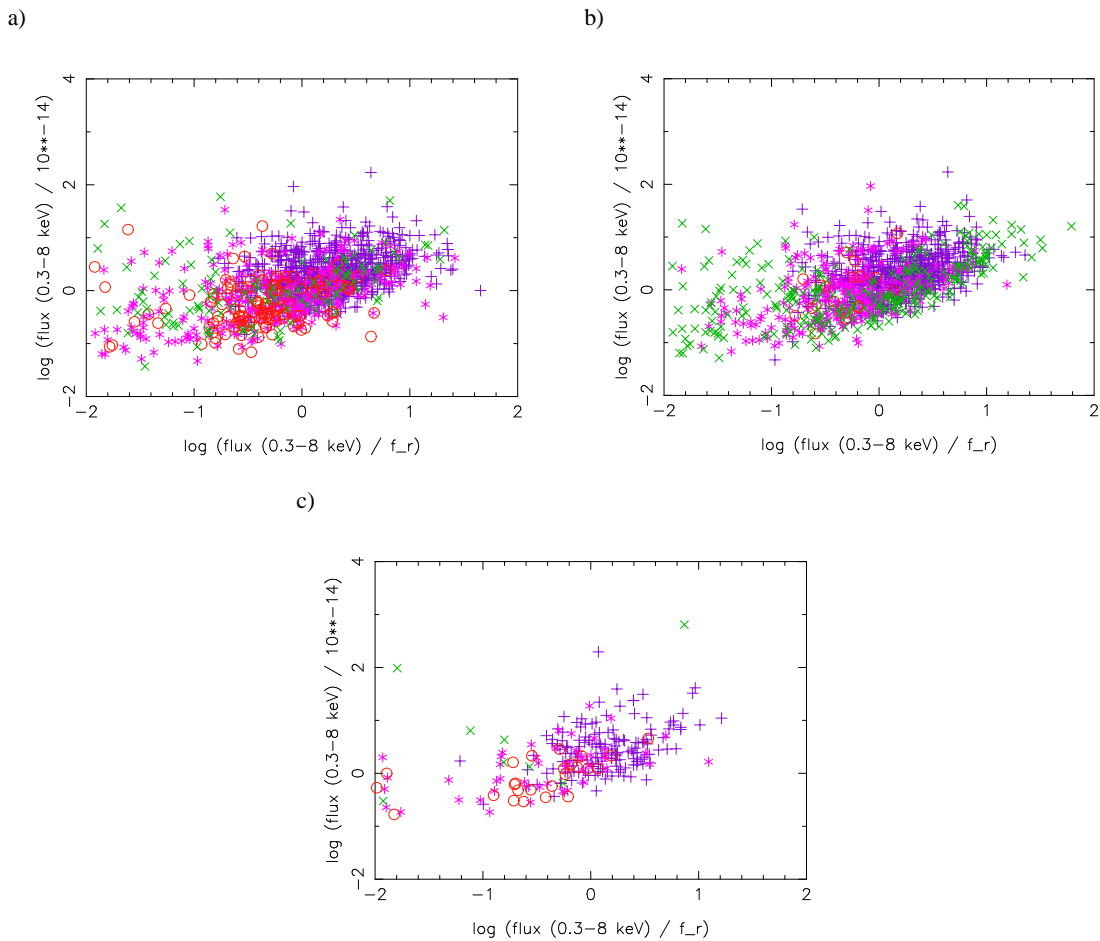

\begin{center}
\begin{tabular}{cc}
a)\includegraphics[height=7cm,angle=-90]{fig16a.eps} & 
b)\includegraphics[height=7cm,angle=-90]{fig16b.eps} \\ 
\\ \multicolumn{2}{c}{c)\includegraphics[height=7cm,angle=-90]{fig16c.eps}} 
\end{tabular}
\caption{Distribution of statistically classified sources in $\log (f_x/f_r)
$-$\log (f_x)$ plane. Plots a) and b) correspond to sets explained in 
Fig. \ref {fig:unidflux}. The same distribution for spectroscopically 
identified is also shown in c). Definition of colors and symbols is the same as 
in Fig. \ref{fig:hf}.}
\label{fig:frfxfx}
\end{center}
\end{figure*}

We have already seen in Fig. \ref {fig:unidflux} the effect of 
under-representation or over representation of a category of objects on the 
classification. Fig. \ref{fig:frfxfx} shows the distribution of corresponding 
statistically classified sources in $\log (f_x/f_r)$-$\log (f_x)$ plane. 
Suppose that from these plots we want to judge about X-ray to optical 
luminosity of 
different category of sources. According to the biased population of 
Fig. \ref{fig:frfxfx}-b, significant number of stars have a very hard spectrum 
i.e. $f_x \gg f_r$. But less biased classification of Fig. \ref{fig:frfxfx}-a 
and spectroscopically classified sources Fig. \ref{fig:frfxfx}-c don't show 
such a population. Similar effects are also visible at low X-ray luminosity, 
low X-ray to optical flux tail where in Fig. \ref{fig:frfxfx}-b this area is 
dominated by stars but Figs. \ref{fig:frfxfx}-a and \ref{fig:frfxfx}-c show 
a significant contribution from NELG and ALG/gal. This has important 
implications for absorption column density in these types of extra-galactic 
objects.

These are examples of misleading conclusions due to incorrect classification. 
Thus before any scientific conclusion be taken from statistical 
classifications, we must first assess their uncertainty and its effect. 
One way to take this into account could be to add an uncertainty to the 
distributions coming from a statistical classification. This is not however a 
simple task. The test of a classification algorithm is based on a small sample 
in which a classification bias can be difficult to detect. When such a 
method is applied to unknown objects, it is very difficult to know whether an 
observed behaviour of data is intrinsic or due to wrong classification. 
Moreover, quantifying the uncertainty of distributions is not simple either. 
We must also add the uncertainty about how representative is the set of 
spectroscopic identifications to classification uncertainty. There are a few 
more issues also to consider: to what extent is the selection of a source for 
spectroscopic follow-up a random process ? How much do observational 
conditions, source location on the sky, visibility to a special telescope, 
etc. bias the selection of sources for follow-up ? Or in general, does the 
set of followed-up fields present a random sampling of the whole sky ? One 
way to answer these questions is to divide the sample, plot separate parts 
and try to find any systematic difference. This is possible only if the size 
of the sample is enough large such that it can be dived to statistically 
significant subsets.

Before concluding this section we want to add few remarks about the general 
purpose 
of statistical identification, in X-ray or other domains of astronomy. As we 
mentioned in this and previous sections, whatever the classification me-thod, 
the input data set leaves its imprint on the statistical identified sources. 
This is quite natural because it is the base of information. Then, the 
question arises: what more we can learn about astronomical issues than what is 
in the input (learning) set ? Probably the most important and unbiased ones 
are issues related to quantities which are not used in the process of 
classification. For instance, large scale spatial correlation of different 
type of objects. More applications can be found with combination of 
observables used 
in the classification - here fluxes - and what is not used in classification 
such as spacial distribution. Note also that redshift is not an independent 
quantity because it is estimated from the knowledge coming from the input set. 

\section{Conclusion}
We have studied a number of statistical methods for automatically 
classifying X-ray selected sources with optical follow-up. We have found 
that a multi-dimensional probability technique that includes maximum 
information ab-out the sources is the best method although it needs a large 
input set. When enough input data is available, with this method BLAGN and 
stars which are dominant sources in the population of the medium and bright 
X-ray sources, are correctly identified in $\sim 90\%$ of cases. Redshift 
distribution can be also recovered with reasonable accuracy. In absence of a 
large input set, a complementary use of various methods can improve 
classification. An interesting point about all the algorithms is that they are 
quite sensitive to the parameter space and their performance can be 
significantly 
improved with a proper choice of parameters. We also studied the effect of 
the input data set and parameter space on the characteristics and relation 
between physical quantities deduced from the statistically classified sources, 
and showed that astronomical conclusions are sensitive both to the input set 
and to the classification method. Nonetheless, because for each class and 
redshift bin we determine a conditional probability, if there is enough 
information about distribution of parameters within a category/redshift, 
classification can have good performance even when the relative contribution 
of categories in the input sample is different from cosmological one.

The general conclusion of this work is that statistical classification 
of X-ray selected sources by some of the me-thods described in this work 
can be sufficiently accurate to be used for astronomical ends. A larger set of 
spectroscopically identified sources is however necessary to improve the 
statistical significance of classifications as well as features detected in 
the statistically classified data. When a significant fraction of ChaMP fields 
are spectroscopically followed up, it would be possible to apply these methods
to a large fraction of the sky and investigate various astronomical 
issues from correlation of Large Scale Structures in X-ray and optical, star 
formation history and evolution of super-massive black holes, etc.

{\bf Acknowledgment} We would like to thank: The ChaMP collaboration for 
putting their valuable data in public access and Kinwah Wu for suggesting the 
use of ChaMP data for this work.

\pagebreak
\oddsidemargin     -1.5cm     
\evensidemargin    -1.5cm     
\onecolumn
{\footnotesize
\begin{center}
\setlongtables
\begin{longtable}{p{26mm}p{10mm}cccllllcccc}
\caption{Summary of statistical classification performance\label{tab:kstest}}\\
\hline 
Method \& Parameter Space (PS) & 
\multicolumn{4}{c}{Input \& Test Data} & \multicolumn{7}{c}{Statistical Classification} & \multirow{3}{5mm}{Fig. ref.}\\
\cline{2-12} & Dep. & Class & Input & Test & {\scriptsize BLAGN} & {\scriptsize NELG} & {\scriptsize ALG} & Star & z & KS & $\chi^2$ & \\
\hline
\endfirsthead
\multicolumn{13}{c}{Summary of statistical classification performance (continue)} \\
\hline 
Method \& Parameter Space (PS) & 
\multicolumn{4}{c}{Input \& Test Data} & \multicolumn{7}{c}{Statistical Classification} & \multirow{3}{5mm}{Fig. ref.} \\
\cline{2-12} & Dep. & Class & Input & Test & {\scriptsize BLAGN} & {\scriptsize NELG} & {\scriptsize ALG} & Star & z & KS & $\chi^2$ &  \\
\hline
\endhead
\endfoot
\multirow{5}{26mm}{Meth.: Distance to clusters. PS: NFXB$^a$} & \multirow{5}{10mm}{Same} & No.Src. & 268 & 268/0(0\%) & 116(77\%) & 22(34\%) & 119(321\%) & 11(73\%) & - & - & - & \multirow{5}{5mm}{-} \\ 
&  & {\scriptsize BLAGN} & 151 & 151 & 84\% & 8\% & 6\% & 2\% & - & - & - & \\ 
&  & {\scriptsize NELG} & 65 & 65 & 27\% & 55\% & 9\% & 9\% & - & - & - & \\ 
&  & ALG & 37 & 37 & 40\% & 35\% & 20\% & 5\% & - & - & - & \\ 
&  & Star & 15 & 15 & 0\% & 27\% & 36.5\% & 36.5\% & - & - & - & \\
\hline
\multirow{5}{26mm}{Meth.: Probability in Optical-X-ray plane. PS: NFOB$^b$} & \multirow{5}{10mm}{Same} & No.Src. & 268 & 268/0(0\%) & 173(115\%) & 60(92\%) & 23(62\%) & 12(80\%) & - & - & - & \multirow{5}{5mm}{-} \\ 
& & {\scriptsize BLAGN} & 151 & 151 & 84\% & 12\% & 3\% & 1\% & - & - & - & \\ 
& & {\scriptsize NELG} & 65 & 65 & 2\% & 53\% & 38\% & 7\% & - & - & - & \\
& & ALG & 37 & 37 & 13\% & 56\% & 26\% & 4\% & - & - & - & \\
& & Star & 15 & 15 & 8\% & 0\% & 25\% & 67\% & - & -  & - & \\
\hline
\multirow{5}{26mm}{Meth.: Low res.spec, $\chi^2$ determined using average 
dispersion. PS: NFXB}  & \multirow{5}{10mm}{Diff.} & 
No.Src. & 143 & 125/9(7\%) & 42(57\%) & 36(129\%) & 37(231\%) & 1(14\%) & - & - & - & \multirow{5}{5mm}{\ref{fig:specred}-b} \\ 
& & {\scriptsize BLAGN} & 77 & 74 & 98\% & 0\% & 0\% & 2\% & 19\% & 42\% & 0.49 &\\ 
& & {\scriptsize NELG} & 37 & 28 & 53\% & 30\% & 14\% & 3\% & 11\% & 32\% & 0.87 &\\ 
& & ALG & 21 & 16 & 32\% & 32\% & 25\% & 11\% & 8\% & 150\% & 4.20 & \\ 
& & Star & 8 & 7 & 0\% & 100\% & 0\% & 0\% & - & - & - &\\
\hline
\multirow{5}{26mm}{Meth.: Low res.spec, $\chi^2$ normalized by template. PS: NFXB} & \multirow{5}{10mm}{Diff.} &
No.Src. & 143 & 125/0(0\%) & 66(89\%) & 26(93\%) & 23(143\%) & 10(142\%) & - & - & - & \multirow{5}{5mm}{-}\\ 
& & {\scriptsize BLAGN} & 77 & 74 & 92\% & 3\% & 3\% & 2\% & 23\% & 18\% & 0.63 & \\ 
& & {\scriptsize NELG} & 37 & 28 & 8\% & 58\% & 23\% & 11\% & 88\% & 14\% & 0.38 & \\
& & ALG & 21 & 16 & 43\% & 35\% & 22\% & 0\% & 83\% & 44\% & 0.56 & \\
& & Star & 8 & 7 & 10\% & 30\% & 30\% & 30\% & - & - & - & \\
\hline
\multirow{5}{26mm}{Meth.: Low res.spec, $\chi^2$ determined using average 
dispersion. PS: NFOB} & \multirow{5}{10mm}{Diff.} &
No.Src. & 143 & 125/5(4\%) & 49(66\%) & 32(114\%) & 8(50\%) & 31(443\%) & - & - & - & \multirow{5}{5mm}{\ref{fig:specred}-c}\\ 
& & {\scriptsize BLAGN} & 77 & 74 & 96\% & 2\% & 0\% & 2\% & 24.5\% & 33\% & 0.64 & \\ 
& & {\scriptsize NELG} & 37 & 28 & 31\% & 44\% & 19\% & 6\% & 90\% & 18\% & 0.55 & \\
& & ALG & 21 & 16 & 12.5\% & 25\% & 62.5\% & 0\% & 87.5\% & 50\% & 0.77 & \\
& & Star & 8 & 7 & 39\% & 32\% & 16\% & 13\% & - & - & - & \\
\hline
\multirow{5}{26mm}{Meth.: Low res.spec, $\chi^2$ normalized by template. PS: NFOB} & \multirow{5}{10mm}{Diff.} &
No.Src. & 143 & 125/1(1\%) & 57(77\%) & 32(111\%) & 31(194\%) & 4(57\%) & - & - & - & \multirow{5}{5mm}{-}\\ 
& & {\scriptsize BLAGN} & 77 & 74 & 96\% & 2\% & 2\% & 0\% & 19\% & 30\% & 0.4 & \\ 
& & {\scriptsize NELG} & 37 & 28 & 28\% & 34\% & 25\% & 12.5\% & 84\% & 14\% & 1.68 & \\
& & ALG & 21 & 16 & 23\% & 48\% & 23\% & 6\% & 97\% & 93\% & 1.34 & \\
& & Star & 8 & 7 & 50\% & 25\% & 0\% & 25\% & - & -  & - & \\
\hline
\multirow{5}{26mm}{Meth.: Low res.spec, $\chi^2$ determined using average 
dispersion, 6 redshift bin. PS: NFOB} & \multirow{5}{10mm}{Diff.} &
No.Src. & 143 & 125/3(2\%) & 61(82\%) & 29(104\%) & 8(50\%) & 24(343\%) & - & - & - & \multirow{5}{5mm}{-}\\ 
& & {\scriptsize BLAGN} & 77 & 74 & 97\% & 2\% & 0\% & 1\% & 43\% & 37\% & 1.21 & \\ 
& & {\scriptsize NELG} & 37 & 28 & 24\% & 49\% & 21\% & 7\% & 90\% & 3\% & 0.06 & \\
& & ALG & 21 & 16 & 12.5\% & 25\%\% & 62.5\% & 0\% & 87.5\% & 50\% & 0.78 & \\
& & Star & 8 & 7 & 21\% & 42\% & 20\% & 17\% & - & - & - & \\
\hline
\multirow{5}{26mm}{Meth.: Maximum Likelihood. PS: NFXB} & \multirow{5}{10mm}{Same} & 
No.Src. & 268 & 268/12(4\%) & 157(104\%) & 51(78\%) & 38(103\%) & 10(67\%) & - & - & - & \multirow{5}{5mm}{-} \\ 
& & {\scriptsize BLAGN} & 151 & 151 & 94\% & 4\% & 1\% & 1\% & 62\% & 3\%  & 0.05 & \\
& & {\scriptsize NELG} & 65 & 65 & 4\% & 92\% & 4\% & 0\% & 98\% & 21\% & 0.07 & \\
& & ALG & 37 & 37 & 3\% & 13\% & 84\% & 0\% & 97\% & 8\% & 0.06 & \\
& & Star & 15 & 15 & 0\% & 0\% & 0\% & 100\% & - & - & - & \\  
\hline
\multirow{5}{26mm}{Meth.: Maximum Likelihood. PS: NFOB} & \multirow{5}{10mm}{Same} & 
No.Src. & 268 & 268/6(2\%) & 151(100\%) & 59(91\%) & 39(105\%) & 13(87\%) & - & - & - & \multirow{5}{5mm}{-} \\
& & {\scriptsize BLAGN} & 151 & 151 & 99\% & 1\% & 0\% & 0\% & 72\% & 5\% & 0.09 & \\
& & {\scriptsize NELG} & 65 & 65 & 2\% & 93\% & 5\% & 0\% & 97\% & 9\% & 0.01 & \\
& & ALG & 37 & 37 & 0\% & 10\% & 87\% & 3\% & 97\% & 5\% & 0.005 & \\
& & Star & 15 & 15 & 0\% & 0\% & 0\% & 100\% & - & - & - & \\ 
\hline
\multirow{5}{26mm}{Meth.: Maximum Likelihood. PS: NFXB} & \multirow{5}{10mm}{Iterr.} &
 No.Src. & 267 & 268/73(27\%) & 145(96\%) & 32(49\%) & 16(43\%) & 1(7\%) & - & - & - & \multirow{5}{5mm}{\ref{fig:mlz}-b} \\ 
& & {\scriptsize BLAGN} & 151 & 151 & 90\% & 6\% & 3\% & 1\% & 34\% & 54\% & 0.15 & \\
& & {\scriptsize NELG} & 65 & 65 & 16\% & 47\% & 34\% & 3\% & 91\% & 50\% & 0.33 & \\
& & ALG & 37 & 37 & 12\% & 44\% & 44\% & 0\% & 81\% & 57\% & 0.47 & \\
& & Star & 15 & 15 & 0\% & 100\% & 0\% & 0\% & - & - & - & \\  
\hline
\multirow{5}{26mm}{Meth.: Maximum Likelihood. PS: NFOB} & \multirow{5}{10mm}{Iterr.} & No.Src. & 267 & 268/72(27\%) & 125(83\%) & 46(71\%) & 22(59\%) & 2(13\%) & - & - & - & \multirow{5}{5mm}{\ref{fig:mlz}-c} \\
& & {\scriptsize BLAGN} & 151 & 151 & 98\% & 2\% & 0\% & 0\% & 29\% & 17\% & 0.17 & \\
& & {\scriptsize NELG} & 65 & 65 & 6\% & 72\% & 22\% & 0\% & 83\% & 29\% & 0.17 & \\
& & ALG & 37 & 37 & 0\% & 41\% & 54.5\% & 4.5\% & 77\% & 41\% & 0.32 & \\
& & Star & 15 & 15 & 0\% & 0\% & 50\% & 50\% & - & - & - & \\ 
\hline
Meth.: Maximum Likelihood. PS: NFXB & Unid. & No.Src. & 268 & 1357/522(38\%) & 588 & 147 & 89 & 11 & - & - & - & \ref{fig:unidmlz}-a \\ 
\hline
Meth.: Maximum Likelihood. PS: NFOB & Unid. & No.Src. & 268 & 1357/639 & 532 & 153 & 23 & 10 & - & - & - & \ref{fig:unidmlz}-b \\ 
\hline
\multirow{5}{26mm}{Meth.: Multi.dim. prob. PS: NFXB} & \multirow{5}{10mm}{Same}& 
No.Src. & 268 & 268/0(0\%) & 155(103\%) & 57(88\%) & 41(111\%) & 15(100\%) & - & - & - & \multirow{5}{5mm}{-} \\
& & {\scriptsize BLAGN} & 151 & 151 & 97\% & 2\% & 1\% & 0\% & 71\% & 13\%  & 0.13 & \\
& & {\scriptsize NELG} & 65 & 65 & 2\% & 98\% & 0\% & 0\% & 100\% & 12\% & 0.017 & \\
& & ALG & 37 & 37 & 0\% & 9\% & 88\% & 0\% & 98\% & 14\% & 0.04 & \\
& & Star & 15 & 15 & 0\% & 0\% & 0\% & 100\% & - & - & - & \\  
\pagebreak
\multirow{5}{26mm}{Meth.: Multi.dim. prob. PS: NFOB} & \multirow{5}{10mm}{Same} & 
No.Src. & 268 & 268/0(0\%) & 152(101\%) & 62(95\%) & 40(108\%) & 14(93\%) & - & - & - & \multirow{5}{5mm}{-} \\
& & {\scriptsize BLAGN} & 151 & 151 & 99\% & 1\% & 0\% & 0\% & 80\% & 7\%  & 0.1 & \\
& & {\scriptsize NELG} & 65 & 65 & 0\% & 98\% & 2\% & 0\% & 98\% & 5\% & 0.003 & \\
& & ALG & 37 & 37 & 0\% & 7.5\% & 90\% & 2.5 & 100\% & 8\% & 0.012 & \\
& & Star & 15 & 15 & 0\% & 0\% & 0\% & 100\% & - & - & - & \\  
\hline
\multirow{5}{26mm}{Meth.: Multi.dim. prob. PS: NFXB} & \multirow{5}{10mm}{Iterr.} & 
No.Src. & 267 & 268/159(59\%) & 90(60\%) & 9(14\%) & 9(24\%) & 0(0\%) & - & - & - & \multirow{5}{5mm}{\ref{fig:probz}-a} \\
& & {\scriptsize BLAGN} & 151 & 151 & 92\% & 6\% & 2\% & 0\% & 24\% & 42\%  & 0.27 & \\
& & {\scriptsize NELG} & 65 & 65 & 22\% & 56\% & 22\% & 0\% & 100\% & 86\% & 0.83 & \\
& & ALG & 37 & 37 & 11\% & 33\% & 56\% & 0\% & 56\% & 76\% & 0.86 & \\
& & Star & 15 & 15 & 0\% & 0\% & 0\% & 0\% & - & - & - & \\  
\hline
\multirow{5}{26mm}{Meth.: Multi.dim. prob. PS: NFOB} & \multirow{5}{10mm}{Iterr.} & 
No.Src. &  267 & 268/170(63\%) & 74(49\%) & 14(26\%) & 8(22\%) & 1(7\%) & - & - & - & \multirow{5}{5mm}{\ref{fig:probz}-b} \\
& & {\scriptsize BLAGN} & 151 & 151 & 99\% & 1\% & 0\% & 0\% & 45\% & 51\% & 0.4 & \\
& & {\scriptsize NELG} & 65 & 65 & 0\% & 79\% & 21\% & 0\% & 93\% & 78\% & 0.72 & \\
& & ALG & 37 & 37 & 0\% & 50\% & 37.5\% & 12.5\% & 100\% & 78\% & 0.92 & \\
& & Star & 15 & 15 & 0\% & 0\% & 100\% & 0\% & - & - & - & \\  
\hline
\multirow{5}{26mm}{Complementary use of 3 algorithms PS: NFXB} & \multirow{5}{10mm}{Iterr.} & 
No.Src. &  267 & 268/64(24\%) & 150(99\%) & 32(49\%) & 21(57\%) & 2(13\%) & - & - & - & \multirow{5}{5mm}{\ref{fig:probmlz}-a}\\
& & {\scriptsize BLAGN} & 151 & 151 & 90\% & 6\% & 3\% & 1\% & 18\% & 3\% & 0.06 & \\
& & {\scriptsize NELG} & 65 & 65 & 19\% & 50\% & 28\% & 3\% & 69\% & 50\% & 0.34 & \\
& & ALG & 37 & 37 & 19\% & 38\% & 43\% & 0\% & 43\% & 43\% & 0.31 & \\
& & Star & 15 & 15 & 50\% & 50\% & 0\% & 0\% & - & - & - & \\  
\hline
\multirow{5}{26mm}{Complementary use of 3 algorithms PS: NFOB} & \multirow{5}{10mm}{Iterr.} & 
No.Src. &  267 & 268/65(24\%) & 129(85\%) & 49(75\%) & 24(65\%) & 4(27\%) & - & - & - & \multirow{5}{5mm}{\ref{fig:probmlz}-b} \\
& & {\scriptsize BLAGN} & 151 & 151 & 98\% & 2\% & 0\% & 0\% & 35\% & 14\% & 0.19 & \\
& & {\scriptsize NELG} & 65 & 65 & 8\% & 70\% & 22\% & 0\% & 69\% & 24\% & 0.14 & \\
& & ALG & 37 & 37 & 4\% & 46\% & 46\% & 4\% & 75\% & 35\% & 0.20 & \\
& & Star & 15 & 15 & 25\% & 0\% & 50\% & 25\% & - & - & - & \\ 
\hline
\multirow{5}{26mm}{Complementary use of 3 algorithms PS: NFXB (only $g$ and $r$ optical bands)} & \multirow{5}{10mm}{Diff.} & 
No.Src. &  143 & 125/7(6\%) & 66(89\%) & 29(104\%) & 21(131\%) & 2(29\%) & - & - & - & \multirow{5}{5mm}{\ref{fig:nou}-a} \\
& & {\scriptsize BLAGN} & 77 & 74 & 85\% & 8\% & 6\% & 1\% & 17\% & 30\% & 0.28 & \\
& & {\scriptsize NELG} & 37 & 28 & 38\% & 41\% & 14\% & 7\% & 62\% & 37\% & 1.04 & \\
& & ALG & 21 & 16 & 29\% & 33\% & 24\% & 14\% & 33\% & 58\% & 0.66 & \\
& & Star & 8 & 7 & 0\% & 0\% & 0\% & 0\% & - & - & - & \\ 
\hline
\multirow{5}{26mm}{Complementary use of 3 algorithms PS: NFOB (only $g$, $r$ and $i$ optical bands)} & \multirow{5}{10mm}{Diff.} & 
No.Src. &  143 & 125/3(2\%) & 65(88\%) & 31(111\%) & 10(63\%) & 16(229\%) & - & - & - & \multirow{5}{5mm}{\ref{fig:nou}-b}\\
& & {\scriptsize BLAGN} & 77 & 74 & 95\% & 3\% & 0\% & 2\% & 23\% & 3\% & 0.06 & \\
& & {\scriptsize NELG} & 37 & 28 & 13\% & 58\% & 26\% & 3\% & 74\% & 50\% & 0.34 & \\
& & ALG & 21 & 16 & 20\% & 40\% & 30\% & 10\% & 10\% & 43\% & 0.31 & \\
& & Star & 8 & 7 & 25\% & 19\% & 31\% & 25\% & - & - & - & \\  
\hline
\multirow{5}{26mm}{Muli.Dim.Prob. PS: NFXB} & \multirow{5}{10mm}{Diff. (Stat.)$^c$} & 
No.Src. & 1331 & 268/119(44\%) & 120(79\%) & 15(23\%) & 12(32\%) & 2(13\%) & - & - & - & \multirow{5}{5mm}{-}\\
& & {\scriptsize BLAGN} & 721 & 151 & 96\% & 3\% & 1\% & 0\% & 64\% & 20\% & 0.27 & \\
& & {\scriptsize NELG} & 332 & 65 & 7\% & 93\% & 0\% & 0\% & 100\% & 77\% &0.67  & \\
& & ALG & 151 & 37 & 0\% & 25\% & 75\% & 0\% & 100\% & 67\% & 0.82 & \\
& & Star & 127 & 15 & 0\% & 0\% & 0\% & 100\% & - & - & - & \\  
\hline
\multirow{5}{26mm}{Complementary use of 3 algorithms PS: NFXB} & \multirow{5}{10mm}{Diff. (Stat.)} & 
No.Src. & 1331 & 268/0(0\%) & 156(103\%) & 53(82\%) & 43(116\%) & 16(107\%) & - & - & - & \multirow{5}{5mm}{\ref{fig:unidflux}-a}\\
& & {\scriptsize BLAGN} & 721 & 151 & 93\% & 6\% & 1\% & 0\% & 52\% & 9\% & 0.30 & \\
& & {\scriptsize NELG} & 332 & 65 & 9\% & 78\% & 9\% & 4\% & 77\% & 18\% &0.07  & \\
& & ALG & 151 & 37 & 2\% & 30\% & 63\% & 5\% & 79\% & 19\% & 0.09 & \\
& & Star & 127 & 15 & 6\% & 6\% & 19\% & 69\% & - & - & - & \\  
\hline
\multirow{5}{26mm}{Muli.Dim.Prob. PS: NFOB} & \multirow{5}{10mm}{Diff. (Stat.)} & 
No.Src. & 1257 & 268/150(56\%) & 99(66\%) & 11(17\%) & 7(19\%) & 1(7\%) & - & - & - & \multirow{5}{5mm}{-} \\
& & {\scriptsize BLAGN} & 676 & 151 & 99\% & 1\% & 0\% & 0\% & 69\% & 33\% & 0.40 & \\
& & {\scriptsize NELG} & 277 & 65 & 0\% & 100\% & 0\% & 0\% & 100\% & 83\% & 1.04 & \\
& & ALG & 22 & 37 & 0\% & 0\% & 100\% & 0\% & 100\% & 81\% & 1.02 & \\
& & Star & 282 & 15 & 0\% & 0\% & 0\% & 100\% & - & - & - & \\ 
\hline
\multirow{5}{26mm}{Complementary use of 3 algorithms PS: NFOB} & \multirow{5}{10mm}{Diff. (Stat.)} & 
No.Src. & 1257 & 268/0(0\%) & 144(95\%) & 52(80\%) & 21(57\%) & 50(333\%) & - & - & - & \multirow{5}{5mm}{\ref{fig:unidflux}-b} \\
& & {\scriptsize BLAGN} & 676 & 151 & 99\% & 1\% & 0\% & 0\% & 41\% & 7\% & 0.33 & \\
& & {\scriptsize NELG} & 277 & 65 & 4\% & 73\% & 19\% & 4\% & 85\% & 20\% & 0.39 & \\
& & ALG & 22 & 37 & 0\% & 28\% & 67\% & 5\% & 71\% & 43\% & 0.88 & \\
& & Star & 282 & 15 & 12\% & 38\% & 26\% & 24\% & - & - & - & \\ 
\hline
\end{longtable}
\end{center}

\begin{description}
\item {$a$} Normalized Fluxes by X-ray Band (NFXB). The parameter space is: 
$\log (f_j/f_B)$ for $j = i, r, g, u$ optical/IR bands, $f_B/ 10^{-14}$ \enerunit, and $\log (f_{x_i}/f_r)$ for $i = S1, S2, H$ X-ray bands.
\item {$b$} Normalized Fluxes by Optical Band (NFOB). The parameter space is: 
$\log (f_j/f_r)$ for $j = z, i, g, u$ optical/IR bands, $f_r/ 10^{-14}$ \enerunit, and $\log (f_{x_i}/f_r)$ for $i = S1, S2, H$ X-ray bands.
\item {$c$} Input data is the set of X-ray sources classified by 
statistical methods. The same method is used both for the classification of 
input and reclassification of spectroscopically identified sources.
\item {$\bullet$} Definition of columns:
\begin{description}
\item Dep.: relation between input and test datasets; Iterr. means when in each 
iterration all available identified sources are used as input except one; 
Stat. means that the input data set comes from statistical classification.
\item Input: Total number and number of each class in the input set.
\item Test: Total number and the number of each class in the test set. The 
number under slash indicates the number(percentage) of sources that couldn't 
be identified.
\item BLAGN, $\cdots$: Detection number and rate $d_i$ ({\it No.Src.} row), 
matrix of reliability $r_i$ (diagonal elements) and contamination rate 
$c_{ij}, i \neq j$, $i, j = $ BLAGN, NELG, ALG/Gal, Star.
\item z: redshift.
\item KS: Kolmogorov-Smirnov (KS) test has been performed by comparing 
accumulative redshift distribution of the test data and their statistical
classification. Percentage is the maximum difference between two distributions.
\item $\chi^2$: The $\chi^2$ has been calculated using equation (\ref {ztest}).
\end{description}
A dash is used when a quantity is not available or relevant.
\end{description}

\end{document}